%% file: Greator.tex
\documentclass[sigconf, nonacm]{acmart}

\usepackage{subcaption}
\usepackage{endnotes,microtype,xspace,graphicx,fancyvrb,multirow} 
\usepackage[linesnumbered, ruled, vlined,resetcount]{algorithm2e} 
\usepackage{algpseudocode}

\usepackage[normalem]{ulem}

\newcommand{\red}[1]{\textcolor{black}{#1}}

\newcommand{\eat}[1]{}

\usepackage{balance}
\input{commands}

\usepackage{comment}

\newcommand{\oursys}{\texttt{Greator}\xspace}
\newcommand{\freshdiskann}{FreshDiskANN\xspace}
\newcommand{\ipdiskann}{IP-DiskANN\xspace}
\newcommand{\delalg}{\texttt{ASNR}\xspace}

\begin{document}

\pagestyle{plain}
 \clearpage
\setcounter{page}{1}

\title{
A Topology-Aware Localized Update Strategy for Graph-Based ANN Index 
}


\settopmatter{authorsperrow=4}

\author{Song Yu}
\affiliation{
  \institution{Northeastern Univ., China}
}
\email{yusong@stumail.neu.edu.cn}

\author{Shengyuan Lin}
\affiliation{
  \institution{Northeastern Univ., China}
}
\email{linsy5847@gmail.com}

\author{Shufeng Gong}
\affiliation{
  \institution{Northeastern Univ., China}
}
\email{gongsf@mail.neu.edu.cn}

\author{Yongqing Xie}
\affiliation{
  \institution{Huawei Tec. Co., Ltd}
}
\email{xieyongqing1@huawei.com}

\author{Ruicheng Liu}
\affiliation{
  \institution{Huawei Tec. Co., Ltd}
}
\email{liuruicheng1@huawei.com}

\author{Yijie Zhou}
\affiliation{
  \institution{Northeastern Univ., China}
}
\email{zhouyijie@stumail.neu.edu.cn}

\author{Ji Sun}
\affiliation{
  \institution{Huawei Tec. Co., Ltd}
}
\email{sunji11@huawei.com}

\author{Yanfeng Zhang}
\affiliation{
  \institution{Northeastern Univ., China}
}
\email{zhangyf@mail.neu.edu.cn}

\author{Guoliang Li}
\affiliation{
  \institution{Tsinghua Univ., China}
}
\email{liguoliang@tsinghua.edu.cn}

\author{Ge Yu}
\affiliation{
  \institution{Northeastern Univ., China}
}
\email{yuge@mail.neu.edu.cn}

\eat{
\author{Lars Th{\o}rv{\"a}ld}
\orcid{0000-0002-1825-0097}
\affiliation{%
  \institution{The Th{\o}rv{\"a}ld Group}
  \streetaddress{1 Th{\o}rv{\"a}ld Circle}
  \city{Hekla}
  \country{Iceland}
}
\email{larst@affiliation.org}

\author{Valerie B\'eranger}
\orcid{0000-0001-5109-3700}
\affiliation{%
  \institution{Inria Paris-Rocquencourt}
  \city{Rocquencourt}
  \country{France}
}
\email{vb@rocquencourt.com}

\author{J\"org von \"Arbach}
\affiliation{%
  \institution{University of T\"ubingen}
  \city{T\"ubingen}
  \country{Germany}
}
\email{jaerbach@uni-tuebingen.edu}
\email{myprivate@email.com}
\email{second@affiliation.mail}

\author{Wang Xiu Ying}
\author{Zhe Zuo}
\affiliation{%
  \institution{East China Normal University}
  \city{Shanghai}
  \country{China}
}
\email{firstname.lastname@ecnu.edu.cn}

\author{Donald Fauntleroy Duck}
\affiliation{%
  \institution{Scientific Writing Academy}
  \city{Duckburg}
  \country{Calisota}
}
\affiliation{%
  \institution{Donald's Second Affiliation}
  \city{City}
  \country{country}
}
\email{donald@swa.edu}
}

\begin{abstract}
The graph-based index has been widely adopted to meet the demand for approximate nearest neighbor search (ANNS) for high-dimensional vectors. 
However, in dynamic scenarios involving frequent vector insertions and deletions, existing systems improve update throughput by adopting a batch update method. 
However, a large batch size leads to significant degradation in search accuracy.

This work aims to improve the performance of graph-based ANNS systems in small-batch update scenarios, 
while maintaining high search efficiency and accuracy.
We identify two key issues in existing batch update systems for small-batch updates. 
First, the system needs to scan the entire index file to identify and update the affected vertices, resulting in excessive unnecessary I/O. Second, updating the affected vertices introduces many new neighbors, frequently triggering neighbor pruning.
To address these issues, we propose 
a topology-aware localized update strategy for graph-based ANN index.
We introduce a lightweight index topology to identify affected vertices efficiently and employ a localized update strategy that modifies only the affected vertices in the index file. 
To mitigate frequent heavy neighbor pruning, we propose a similar neighbor replacement strategy, which connects the affected vertices to only a small number (typically one) of the most similar outgoing neighbors of the deleted vertex during repair.
Based on extensive experiments on real-world datasets, 
our update strategy achieves \red{2.47$\times$-6.45$\times$} higher update throughput than the state-of-the-art system \freshdiskann while maintaining high search efficiency and accuracy.

\end{abstract}

\maketitle

\input{1-intro}

\input{2-bg}

\input{3-overview}

\input{4-1-IO}
\input{4-2-compute}

\input{4-3-impl}

\input{5-eval}

\input{6-related}

\input{7-concl}

\eat{
\begin{acks}
 This work was supported by the [...] Research Fund of [...] (Number [...]). Additional funding was provided by [...] and [...]. We also thank [...] for contributing [...].
\end{acks}
}


\bibliographystyle{ACM-Reference-Format}
\bibliography{sample}

\end{document}

%% file: commands.tex
\usepackage{latexsym}
\usepackage{amsfonts}
\usepackage{amsmath}
\usepackage{amssymb}
\usepackage{color}
\usepackage{epsfig}
\usepackage{xspace}
\usepackage{graphicx}
\usepackage{cleveref}
\usepackage{balance}
\usepackage{hhline}
\usepackage{float}
\usepackage{xcolor}

\usepackage{epsfig}
\usepackage{multirow}
\usepackage{url}

\usepackage{enumitem}
\setlist{topsep=0pt,noitemsep} \setitemize[1]{label=$\circ$}

\newcommand{\ei}{\end{itemize}\vspace{1ex}}

\newcommand{\ee}{\end{enumerate}\vspace{1ex}}
\newcommand{\beqn}{\begin{eqnarray}}
\newcommand{\eeqn}{\end{eqnarray}}

\newcommand{\stitle}[1]{\vspace{1.2ex}\noindent{\bf #1}}
\newcommand{\etitle}[1]{\vspace{1.2ex}\noindent{\em\uline{#1}}}

\newcommand{\ie}{i.e.,\xspace}
\newcommand{\eg}{e.g.,\xspace}

\renewcommand{\smallskip}{\vspace{0.6ex}}

\newcounter{ccc}

\newcommand{\eop}{\hspace*{\fill}\mbox{$\Box$}}     
\newcounter{example}
\renewcommand{\theexample}{\arabic{example}}
\newenvironment{example}{
        \vspace{1.2ex}
        \refstepcounter{example}
        {\noindent\bf Example \theexample:}}{
        \eop\vspace{1.2ex}}

\newlist{myitemize}{itemize}{3}
\setlist[myitemize,1]{label=$\circ$,leftmargin=3.8ex}
\setlist[myitemize,2]{label=$\bullet$,leftmargin=3.8ex}
\setlist[myitemize,3]{label=$\diamond$, leftmargin=3.2ex}

\newcommand{\mei}{\end{myitemize}\vspace{0.6ex}}

\usepackage{tikz}

%% file: 1-intro.tex
\section{Introduction}
\label{sec:intro}

Approximate nearest neighbor search (ANNS)
for high-dimensional vectors has become a critical component in modern data-driven applications, with widespread usage in information retrieval \cite{AsaiMZC23,WilliamsLKWSG14,MohoneyPCMIMPR23,SanthanamKSPZ22}, recommendation systems \cite{MengDYCLGLC20,SarwarKKR01,CovingtonAS16,OkuraTOT17}, and large language models (LLM) \cite{abs-2304-11060,abs-2312-10997,LewisPPPKGKLYR020,abs-2202-01110,jing2024large,bang2023gptcache,abs-2411-05276,VELO,abs-2502-03771}. 
To accelerate ANNS while maintaining high search accuracy, researchers have developed numerous index techniques.
These indexes can be categorized into two types: graph-based indexes \cite{HNSW,NSG,HM-ANN,DiskANN,SymphonyQG,FreshDiskANN,Starling,Filtered-DiskANN} and partition-based (clustering-based) indexes \cite{R-Tree, Rstar-Tree,IVFADC,LSH,SPANN,spfresh}, where graph-based indexes have gained widespread attention due to their superior trade-off between search efficiency and accuracy in high-dimensional scenarios~\cite{Starling,SymphonyQG,WangXY021,ANN-Benchmarks,LiZSWLZL20,azizi2025graph}. Many vector search systems have adopted graph-based indexes to enable efficient retrieval~\cite{Starling,DiskANN,OOD-DiskANN,FreshDiskANN,ParlayANN,SPTAG}. 
However, for large amounts of high-dimensional vectors, the graph-based index tends to be very large, often exceeding the memory capacity of ordinary commodity PCs \cite{Starling}.
Therefore, existing systems commonly store indexes on disk~\cite{DiskANN,OOD-DiskANN,FreshDiskANN,Starling}, enabling cost-effective scalability.

With the widespread adoption of ANNS, an increasing number of applications require these systems to support large-scale dynamic updates (including insertions/deletions) and vector queries. To meet these demands, modern ANNS systems need to handle billions of vector updates daily while maintaining low search latency and high search accuracy \cite{spfresh}. This is particularly crucial for many AI applications built on these systems, where real-time updates are essential, such as in emails and chat histories, all of which are converted into embedding vectors to retrieve the most relevant fragments as new prompts.

\begin{figure} 
\centering
    \begin{minipage}{0.48\linewidth}
        \subfloat{\includegraphics[width = 1\linewidth]
            {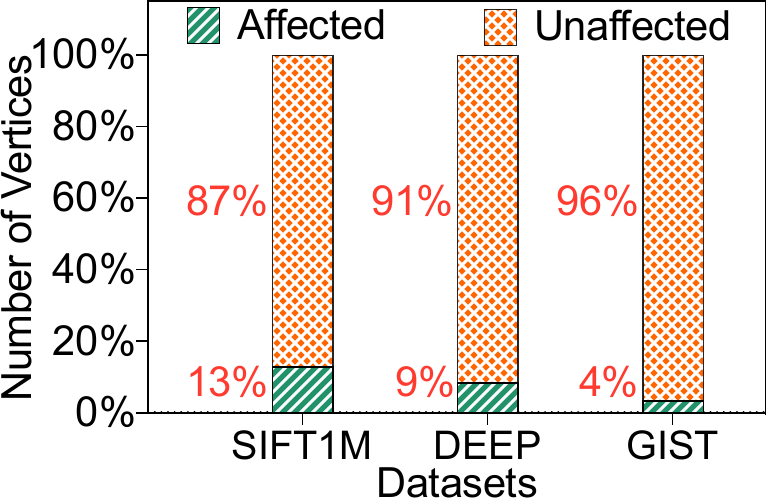}\label{fig:necessary_vertex} \hspace{0in}}
        \vspace{-0.12in}
        \captionsetup{font=small}
        \caption{
        The ratio of affected vertices to unaffected vertices in the index.
        }
        \label{fig:effctive_page_percentage}
    \end{minipage}
    \ \ \
        \begin{minipage}{0.48\linewidth}
        \subfloat{\includegraphics[width = 1\linewidth]
            {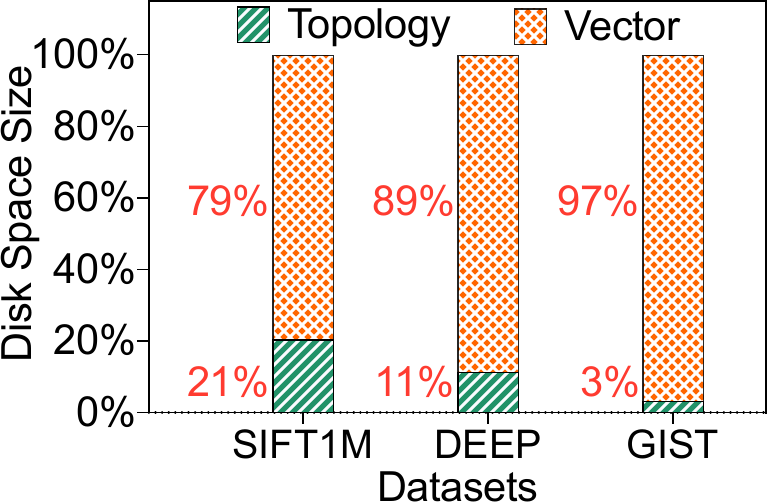}\label{fig:vector_topo} \hspace{0in}}
        \vspace{-0.12in}
        \captionsetup{font=small}
        \caption{The ratio of disk space occupied by vectors and graph topology in the index.
        }
        \label{fig:vector_topo}
    \end{minipage}

  \vspace{-0.15in}
\end{figure}

\stitle{Motivation.} 
Although existing graph-based vector search systems have demonstrated excellent performance in search efficiency, accuracy and scalability, they experience significant performance bottlenecks when handling 
vector updates
with frequent insertions and deletions.
Existing systems update the graph-based index through batch processing to amortize the overhead of single update operations \cite{FreshDiskANN,Milvus,AnalyticDB-V,pgvector}. 
However, we find that batch update strategies face a trade-off between update efficiency and search accuracy. 
We evaluate \freshdiskann\cite{FreshDiskANN}, a state-of-the-art batch update ANNS system on disk, with different batch sizes on the GIST dataset (detailed in Table \ref{tab:dataset}).
We find that\eat{when the batch size increases from \red{0.1\%} to \red{8\%} of each dataset, the update throughput increases by \red{8.78-10.51}$\times$, but the search accuracy (\ie recall) drops by \red{7.75-8.39\%} (see Figure \ref{fig:vary_batch_size} in Section \ref{sec:expr:vary_batch_size}).}
as the batch size increases, the update throughput improves significantly, but the search accuracy (\ie recall) decreases. 
Specifically, when the batch size increases from \red{0.1\%} to \red{8\%} of the dataset, the update throughput improves by \red{8.78$\times$}, while the search accuracy drops by \red{8.39\%} (detailed in Figure \ref{fig:vary_batch_size} of Section \ref{sec:expr:vary_batch_size}). 
This implies that 
improving update performance by increasing the batch size will sacrifice the search accuracy of the index. \textit{In light of this, is it possible to enhance the throughput of graph index updates, even with a small update batch?}

\stitle{Our goal}. In this work, we aim to improve the update performance of graph-based index under small-batch updates while maintaining high search efficiency and accuracy.

\stitle{Drawbacks of graph updating}. 
There are two drawbacks in existing methods under small-batch updates:

\etitle{(1) 
Inefficient 
disk I/O}. 
Existing systems primarily suffer from substantial unnecessary I/O for two reasons. 
First, the system needs to scan the graph topology of the index to identify and repair all affected vertices whose neighbors have been deleted. 
However, in small-batch update scenarios, the proportion of affected vertices is very small. 
For example, as shown in Figure \ref{fig:effctive_page_percentage}, when \red{0.1\%} of the vertices are inserted and deleted in three real-world datasets, only \red{4\%-13\%} of the vertices are affected.
Despite the small proportion of affected vertices, existing methods update them by traversing the entire index file, leading to substantial unnecessary I/O (e.g., \red{96\%} of the I/O is unnecessary in the GIST dataset).

Second, to optimize search performance, existing indexes usually store both the vector and neighbors of each vertex together, enabling a single I/O request to retrieve them \cite{DiskANN,Starling,FreshDiskANN,Filtered-DiskANN,OOD-DiskANN}.
This coupled storage method results in excessive unnecessary I/O when scanning the graph index to find the affected vertices by the deleted vertex, 
since reading vectors is not required. 
This issue is particularly severe in high-dimensional scenarios, where vector data is typically much larger than the graph topology. 
As shown in Figure \ref{fig:vector_topo}, on three real-world datasets—SIFT1M (128 dimensions), DEEP (256 dimensions), and GIST (960 dimensions) (detailed in Table \ref{tab:dataset}), the graph topology with \red{32} neighbors accounts for only \red{3\%-21\%} of the total index size, indicating that \red{79\%-97\%} of the data read is unnecessary.

\etitle{(2)
Frequent 
heavy neighbors pruning}. 
When a vertex's neighbors are deleted, existing methods repair the affected vertex to maintain the graph's navigability by connecting it to all neighbors of the deleted vertices, thereby compensating for the lost connectivity.
However, this approach significantly increases the number of neighbors of affected vertices, 
which leads to more edges being traversed during the search, thereby increasing search latency. 
To mitigate this issue, existing methods prune neighbors to keep their number below a predefined limit. However, 
pruning requires computing similarities (distances) between all pairs of neighbors, which severely degrades update performance, especially in high-dimensional scenarios.

\stitle{Our Solution}.
To overcome the drawbacks of index updating with small-batch updates,
we design a topology-aware localized update strategy for graph-based ANNS index on high-dimensional streaming vectors.
The core idea is to leverage the localized nature of small-batch updates, where only a few vertices are affected and their impact on the graph structure is localized. This allows for 
local updates to the index file, reducing unnecessary I/O and computation.
Specifically, our strategy incorporates several key designs to achieve fast updates.
(1) \textbf{Topology-Aware Affected Vertices Identification}. We design a topology-aware ANNS index that includes a query index and a lightweight topology. 
The query index is used for fast vector search, while the lightweight topology is employed to quickly identify affected vertices by avoiding unnecessary I/O caused by reading large amounts of vector data.
(2) \textbf{Localized Update}. We design a localized update method that updates the index by only reading the pages containing the affected vertices, thereby avoiding unnecessary I/O. 
(3) \textbf{Similar Neighbor  Replacement}. 
We design a similar neighbor replacement method 
which connects the affected vertices to only a small number (typically one) of the most similar
outgoing neighbors of the deleted vertex during the repair.
This localized repair method prevents high-cost pruning caused by adding too many neighbors.

In summary,
we make the following contributions.
\begin{itemize}[leftmargin=0.2in]

\item[$\bullet$] 
We propose a topology-aware index that efficiently identifies the affected vertices through a lightweight topology (Section \ref{sec:io:index}). Additionally, we provide a localized update method that minimizes I/O to update the index, thereby avoiding the overhead caused by unnecessary I/O (Section \ref{sec:io:update}).

\item[$\bullet$]
We propose a similar neighbor replacement method that prevents frequently triggering costly neighbor pruning, thereby reducing the expensive vector distance computations (Section \ref{sec:cmp}).

\item[$\bullet$] 
Based on \eat{all }the designs presented in this paper, we implement a \underline{gr}aph-based str\underline{ea}ming vec\underline{tor} ANNS system, \oursys, that supports localized incremental updates (Section \ref{sec:impl}), and we conduct a comprehensive evaluation to verify its efficiency (Section \ref{sec:expr}). 
Experimental results show that on real-world datasets,
\oursys achieves \red{2.47$\times$-6.45$\times$} higher update throughput than the state-of-the-art system \freshdiskann while maintaining high search efficiency and accuracy.

\end{itemize}

%% file: 2-bg.tex
\section{
Background and MOTIVATION}
\label{sec:bg}

\begin{figure*}[tbp]
    \centering
    \includegraphics[width=0.95\linewidth]{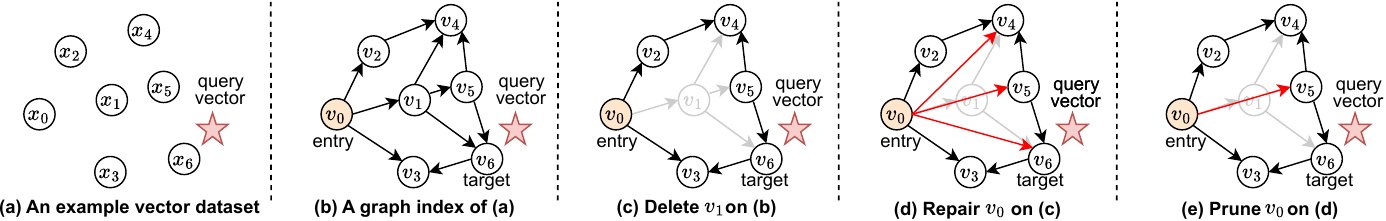}
    \vspace{-0.08in}
    \caption{\eat{Illustrate a}An example of a graph-based vector index and its update process. 
    Each vertex has a maximum neighbor limit of $R=3$. The red star represents the query vector, $v_0$ represents the entry\eat{ vertex} of the index, and $v_6$ ($x_6$) represents the nearest target vertex (vector) for the query.
    }
    \vspace{-0.05in}
    \label{fig:graph_index_eg}
\end{figure*}

This section first introduces the background of graph-based approximate nearest neighbor search (ANNS), followed by a detailed explanation of the update methods for graph-based indexes on disk.

\subsection{Graph-Based ANNS Index}
\label{sec:bg:graph_index}

Let $X=\{x_1, x_2, ..., x_n\}$ denote a vector dataset of $n$ vectors, where each vector $x_i \in \mathbb{R}^{d}$ represents a $d$-dimensional vector.
The distance between two vectors, $x_p \in \mathbb{R}^{d}$ and $x_q \in \mathbb{R}^{d}$, is represented as $dist(x_p, x_q)$. The distance function can be the Euclidean distance, cosine similarity, or other metrics.

\stitle{Approximate Nearest Neighbors Search (ANNS)}. 
Given a vector dataset $X$ and a query vector $x_q \in \mathbb{R}^{d}$, the goal of approximate nearest neighbor search (ANNS) is to retrieve a set $R_{knn}$ of $k$ vectors from $X$ that are closest to $x_q$, note that the retrieved results are not guaranteed to be optimal.

Typically, the accuracy of the search result $R_{knn}$ is evaluated using the \textit{recall}, defined as $k\text{-recall}@k = \frac{| R_{knn} \cap R_{exact}|}{k}$, where $R_{exact}$ is the ground-truth set of the $k$ closest vectors to $x_q$ from $X$. The goal of ANNS is always to maximize the recall while retrieving results as quickly as possible, leading to a trade-off between recall and latency.

\stitle{Graph-Based Vector Index}. 
Given a vector dataset $X$ and a graph-based vector index $G = (V, E)$ of $X$, where $V$ is the set of vertices with a size equal to $|X|$. Each vertex in $V$ corresponds to a vector in $X$.
Specifically, $x_p$ denotes the vector data associated with vertex $p$, and $dist(p, q)$ represents the distance between two vertices $p$ and $q$. 
$E$ is the set of edges constructed by a specific index algorithm based on 
vector similarity. 
$N_{out}(p)$ and $N_{in}(p)$ denote the outgoing neighbor set and incoming neighbor set of $p$, defined as $N_{out}(p) = \{v|edge(p, v) \in E\}$ and $N_{in}(p) = \{v | edge(v, p) \in E\}$.
Figure~\ref{fig:graph_index_eg}a illustrates an example vector dataset $X$, while Figure~\ref{fig:graph_index_eg}b presents its corresponding graph-based index.

Existing graph indexes typically only maintain unidirectional graphs with outgoing neighbors, especially in disk-based implementations.
This is because, in graph-based indexes \cite{HNSW,DiskANN,NSG,Starling,OOD-DiskANN}, the number of outgoing neighbors for each vertex is roughly the same and does not exceed a threshold $R$, allowing fixed space allocation for outgoing neighbors and enabling independent updates. In contrast, the number of incoming neighbors varies significantly, and allocating fixed space for them leads to storage waste. 
On the other hand, compact storage requires adjusting the storage locations of subsequent vertices as the number of neighbors increases, \eat{which incurs overhead that is unacceptable}{incurring unacceptable overhead} \cite{lsmgraph,livegraph,gastcoco}. Moreover, ensuring consistency in bidirectional graphs during concurrent updates significantly \eat{affects}degrades index update performance.

\stitle{ANNS with Graph-Based Index}. 
Graph-based ANNS typically employs the beam search algorithm~\cite{DiskANN}, a variant of best-first search. Specifically, the algorithm initializes a priority queue with a predefined entry\eat{ point}, which can be randomly selected or determined using a heuristic method. 
In each iteration, the algorithm performs a greedy search by extracting the $W$ vertices closest to the query vector from the queue, evaluating their neighbors, and inserting eligible candidates back into the queue. 
This process continues until a stopping criterion is met, such as reaching a predefined search limit $L$. By adjusting $L$, beam search achieves a trade-off between search accuracy and efficiency.

\subsection{Disk-Based Graph Index Updates}
\label{sec:bg:existing_work}

As introduced in Section \ref{sec:intro}, modern applications increasingly require dynamic updates (insertion/deletion) and queries for vector data \cite{AnalyticDB-V,Ada-IVF,spfresh,FreshDiskANN,Milvus}.
In these applications, the index needs to promptly respond to vector updates to ensure search accuracy and efficiency. 
For example, as shown in Figure~\ref{fig:graph_index_eg}c, when vertex $v_1$ (corresponding to vector $x_1$) is deleted, the existing index structure fails to return the target search result (\ie vertex $v_6$) because there is no path from the entry (\ie $v_0$) to reach it.
Therefore, the index structure needs to be updated accordingly. 
Figure~\ref{fig:graph_index_eg}d illustrates a typical repair process used in existing methods \cite{FreshDiskANN}, where the incoming neighbor $v_0$ of the deleted vertex $v_1$ is directly connected to \eat{its}$v_1$'s outgoing neighbor set $\{v_4, v_5, v_6\}$. 
\eat{Then}Subsequently, the neighbors of $v_0$ are pruned to \eat{stay within}satisfy a predefined limit $R$ (set to $R=3$ in the figure). 
The final updated index is shown in Figure~\ref{fig:graph_index_eg}e, where the modified index structure allows ANNS to efficiently and accurately retrieve the target vertex $v_6$.

However, efficiently updating indexes while maintaining search accuracy remains a significant challenge.
As shown in Figure~\ref{fig:graph_index_eg}d, when a vertex is deleted, the affected incoming neighbors must be repaired to maintain the navigability of the index. 
\eat{However, a}As discussed in Section~\ref{sec:bg:graph_index}, existing graph-based indexes typically maintain a directed topology that only stores outgoing neighbors. Consequently, to identify and repair all affected vertices (\ie incoming neighbors of deleted vertices), existing methods must traverse the entire index. 
\eat{Moreover}Furthermore, since graph indexes in existing systems are often stored on disk to support large-scale vector data, the update overhead is further amplified. 

To amortize the update cost, existing methods generally adopt a batch update mechanism \cite{Milvus,FreshDiskANN,AnalyticDB-V,pgvector}. 
Next, we detail the specific steps, which consist of three phases: deletion, insertion, and patch.

\etitle{Deletion Phase.} 
This phase repairs the neighbors of all vertices affected by the deleted vertex $p$ (i.e., the incoming neighbors of $p$).
However, since graph-based indexes do not store the incoming neighbors\eat{ (i.e., $N_{in}(p)$)} of each vertex, they cannot directly locate $N_{in}(p)$\eat{by the deleted vertex $p$}. 
As a result, 
it is necessary to load the vertices and their neighbors block by block from the disk, execute Algorithm \ref{alg:delete} \cite{DiskANN} on the vertices within each block, and write the modified blocks to a temporary intermediate index file on the disk.

As shown in Algorithm \ref{alg:delete}, for each affected vertex $p$, a new candidate neighbor set $\mathcal{C}$ is constructed. 
The vertices in $N_{out}(p)$ that are not deleted are added to $\mathcal{C}$. 
Additionally, for each deleted vertex $v$ in $N_{out}(p)$, the vertices in $N_{out}(v)$ that are not deleted are also added to $\mathcal{C}$.  
Next, it triggers neighbor pruning, \ie the RobustPrune algorithm~\cite{FreshDiskANN}, to control the size of $\mathcal{C}$. 
The pruned result is then assigned as the new neighbor set of $p$. 
It is worth noting that the RobustPrune algorithm \cite{DiskANN} has a computational complexity of $O(|\mathcal{C}|^2 \times d)$, where $d$ is the dimension of \eat{the}each vector.

\begin{algorithm}
\small
\caption{Delete($L_D, R, \alpha$) \cite{FreshDiskANN} }
\label{alg:delete}
\KwIn{Graph $G(V, E)$ with $|V| = n$, set of vertices to be deleted $L_D$}
\KwOut{Graph on nodes $V'$ where $V' = V \setminus L_D$}

\Begin{
    \ForEach{$p \in V \setminus L_D$ \textbf{s.t.} $N_\text{out}(p) \cap L_D \neq \emptyset$}{
        $\mathcal{D} \gets N_\text{out}(p) \cap L_D$\;
        $\mathcal{C} \gets N_\text{out}(p) \setminus \mathcal{D}$ \tcp*[r]{initialize candidate list}
        \ForEach{$v \in \mathcal{D}$}{
            $\mathcal{C} \gets \mathcal{C} \cup (N_\text{out}(v) \setminus \mathcal{D})$\;
        }
        $N_\text{out}(p) \gets \text{RobustPrune}(p, \mathcal{C}, \alpha, R)$\;

    }
}
\end{algorithm}

\etitle{Insertion Phase.} 
For each inserted vertex $p$, a greedy search (\ie $GreedySearch$ \cite{FreshDiskANN}) is performed on the graph, which requires random reads from the index file. Based on the search results, the neighbors $N_{out}(p)$ of the vertex are constructed. \eat{Additionally}Furthermore, for all outgoing neighbors of $p$ (\ie $N_{out}(p)$), a reverse edge to $p$ needs to be added, \ie $\{edge(p^{'}, p)| p \in N_{out}(p)\}$. 
However, immediately inserting the vector, neighbors, and all reverse edges of vertex $p$ causes random writes. To avoid this, this phase stores these updates in a temporary in-memory structure $\Delta$ to merge writes to the same page, which is processed by the next phase.

\etitle{Patch Phase.} This phase applies the updates stored in $\Delta$ from the insertion phase to the temporary index file and generates a new index file. Specifically, it sequentially retrieves all vertices $p$ from the temporary index file on the SSD in blocks, adds the outgoing edges of each vertex $p$ from $\Delta$, and checks whether the new degree $|N_{\text{out}}(p) \cup \Delta(p)|$ exceeds a given threshold (\eg $R$). 
If the threshold is exceeded, \eat{it}the algorithm triggers a neighbor pruning using the $RobustPrune$ algorithm \cite{FreshDiskANN}. 
Finally, all updated blocks are written to the new index file on the SSD.

In conclusion, existing batch update methods identify and repair affected vertices by performing two index file scans during the deletion and patch phases, leading to significant I/O overhead. Furthermore, during neighbor updates, existing repair algorithms introduce excessive neighbors, frequently triggering neighbor pruning and further increasing computational overhead.

%% file: 3-overview.tex
\section{Overview}
\label{sec:overview}

\begin{figure}[tbp]
    \centering
    \includegraphics[width=1\linewidth]{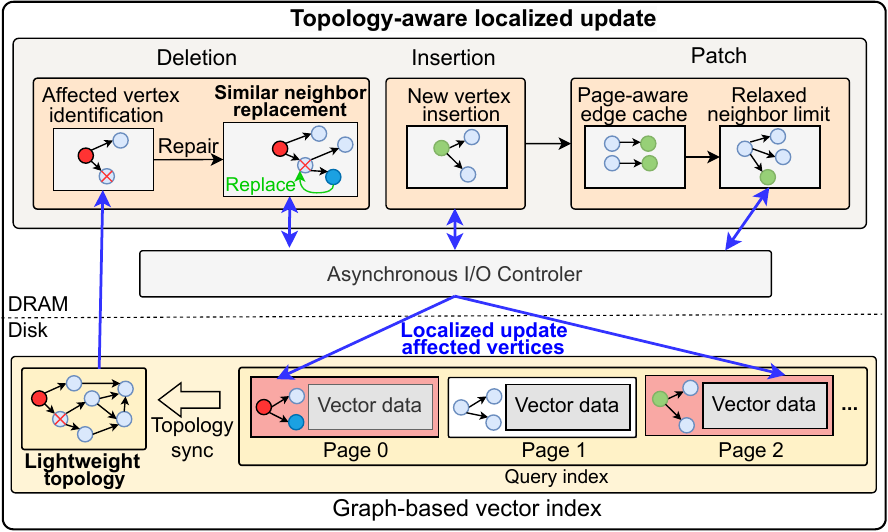}
    \vspace{-0.15in}
    \caption{The update workflow of our topology-aware localized update strategy. The crossed-out circles represent deleted vertices, the red circles indicate vertices affected by the deletion, and the green circles denote newly inserted vertices. 
    The red pages (\eg page 0 and page 2) represent the affected pages that need to be updated.
    }
    \label{fig:workflow}
    \vspace{-0.15in}
\end{figure}

We present a topology-aware localized update strategy that enables fast updates of graph-based index structures in small-batch update scenarios.
Our update strategy primarily enhances update performance through the following core designs. 
To reduce I/O overhead, 
(1) we introduce a topology-aware ANNS index, which includes a query index and a lightweight topology. The query index is used for fast vector search, while the \textbf{lightweight topology} efficiently identifies affected vertices, avoiding unnecessary I/O caused by scanning the entire index.
(2) We design a \textbf{localized update method}, updating the index locally based only on the affected vertices, thereby avoiding modifications to the entire index file. 
To reduce computational overhead, 
(3) we propose a \textbf{similar neighbor replacement method}, which connects the affected vertices to only a small number of the most similar outgoing neighbors of the deleted vertex during repair. This method adapts to the current number of neighbors to select the appropriate number of replacements and ensures that the neighbor limit is not exceeded, thus reducing frequent and heavy neighbor pruning.

\stitle{Update Workflow}. 
After accumulating a small batch of updates, our strategy efficiently handles the batch and merges the updates to the graph-based vector index file on disk. 
As shown in Figure~\ref{fig:workflow}, this update workflow can be divided into the following three stages.
1) \textit{\textbf{Deletion Phase}.}
This phase begins by processing the deletion vectors. 
It scans a lightweight topology significantly smaller than the query index file to quickly locate the incoming neighbors affected by each deleted vertex. 
Next, it loads only the affected pages from the query index file to reduce unnecessary I/O through localized updates. It then repairs the neighbors of these vertices using the newly designed similar neighbor replacement method, which significantly reduces the frequency of neighbor pruning.
Finally, the updated pages are written back to the query index file.  
2) \textit{\textbf{Insertion Phase}.} 
This phase processes the insertion vectors. 
Each insertion vector is treated as a new vertex, which undergoes an approximate search on the query index to construct its neighbors based on the search results. 
Based on the search results, these new vertices, along with their neighbors and vector data, are written to the query index file. 
Simultaneously, the reverse edges of the outgoing neighbors of the newly inserted vertices are cached in a page-aware cache structure $\Delta G$.  
3) \textit{\textbf{Patch Phase}.} 
Based on the information in $\Delta G$, this phase merges the reverse edges to be added for the same vertex. It then reads only the pages corresponding to the affected vertices, updates their neighbors using a relaxed neighbor constraint, and writes the updated pages back to the index file.  

%% file: 4-1-IO.tex
\section{Topology-Aware ANNS Index with Localized Update}
\label{sec:design}

In this section, 
we propose a novel index that can be updated via localized updates with minimal disk I/O.
Section~\ref{sec:io:index} provides a detailed explanation of the update-friendly design of this index, while Section~\ref{sec:io:update} presents the page-level localized update strategy tailored for this index.

\subsection{A Topology-Aware ANNS Index
}
\label{sec:io:index}

As discussed in Section \ref{sec:bg:existing_work}, existing methods face a common dilemma when updating indexes: they require two scans of the index to identify and repair the affected vertices. The root cause lies in the fact that these methods primarily optimize for search efficiency and accuracy when designing the index but neglect update performance. To address both search and update performance, we design a new index structure. The index consists of two parts: a query index to optimize search efficiency and a lightweight topology to optimize update performance. The design details are explained below.

\stitle{Query Index}.
To meet the efficiency and accuracy requirements of vector search, we adopt the state-of-the-art disk-based graph index methods (\ie DiskANN \cite{DiskANN}) to construct a static query index. 
In the storage format of the query index on disk, each vertex's outgoing neighbors and vector data are stored together in a compact manner.
This design allows both neighbor data and vector data to be obtained during the search process, thereby reducing I/O operations.

Although the query index satisfies the efficiency and accuracy requirements for vector search like DiskANN, it still faces the same update problem as existing methods: it cannot quickly identify the vertices affected by deletions because it does not store the incoming neighbors of vertices. 
As discussed in Section \ref{sec:intro}, existing methods identify affected vertices by scanning the entire index file. However, since the index file contains a large amount of vector data, this scan results in \eat{a significant number of }significant unnecessary I/O reads. To address this issue, we design a lightweight topology separately, which replaces the query index to quickly identify the affected vertices.

\stitle{Lightweight Topology}.
We design a lightweight topology that stores only the neighbors of each vertex. Its content is consistent with the neighbors in the query index. By eliminating the vector data, the size of the lightweight topology is typically much smaller than that of the query index file. As shown in Section \ref{sec:intro}, on real GIST vector datasets, the graph topology data accounts for only \red{3\%} of the index file. Based on this design, the update process only requires scanning the lightweight topology file to quickly identify the affected vertices without scanning the query index, thus avoiding the unnecessary I/O found in existing methods.

\stitle{Index Consistency}.
Since both the query index and the lightweight topology store the graph's topological information, it is necessary to ensure consistency between the two during updates. Note that the lightweight topology is only used to identify the affected vertices and is not utilized during the search process. Therefore, we adopt a lazy synchronization update strategy: the query index is updated first, followed by synchronizing the updated structural information to the lightweight topology in the background. This strategy has two key advantages: (1) It only synchronizes the data that has changed. 
When the neighbors of vertex $p$ in the query index are updated, only the information for $p$ in the lightweight topology needs to be synchronized; (2) the background synchronization has minimal impact on the performance of updates.

\subsection{Page-Level Localized Updates}
\label{sec:io:update}

As described in Section \ref{sec:bg:existing_work}, existing systems update affected vertices by sequentially scanning the entire index file. This approach is reasonable in large-batch update scenarios, as most of the sequentially loaded data requires modification, and the full sequential scanning utilizes the high bandwidth of sequential I/O. 
However, in small-batch update scenarios, the data loaded contains a large number of unaffected vertices.
As shown in Figure \ref{fig:effctive_page_percentage}, existing systems read a large number of unnecessary vertices, resulting in significant unnecessary I/O overhead. 

\stitle{Page-Level Update}.
To eliminate the issue of unnecessary I/O in existing methods, we adopt a fine-grained page-level localized update method, updating only the pages that contain affected vertices. Specifically, it follows the three-phase batch update process: deletion, insertion, and patch.

\etitle{Deletion}.
For each deleted vertex $p$, our method removes $p$ only from the mapping $Local\_Map$, which records the association between vertices and their locations in the index file. It then adds the freed location to a recycling queue $Free\_Q$.
After that, it utilizes the lightweight topology to quickly scan and identify the set of all affected vertices (\ie those whose neighbors include the deleted vertex), denoted as $V_{affect}$.
For an affected vertex $q \in V_{affect}$, our method first locates its corresponding storage page in the index file using $Local\_Map$. It then loads the relevant data page via an efficient asynchronous I/O controller interface (detailed in Section \ref{sec:impl}). 
Once the loading is complete, a lightweight neighbor repair algorithm (\delalg, detailed in Section \ref{sec:cmp:delete}) updates the outgoing neighbor set $N_{out}(q)$. 
Finally, the modified data page is written back to the index file.

\etitle{Insertion}.
For each inserted vertex $p$, our method constructs the outgoing neighbor set $N_{out}(p)$ using the same method as existing systems (detailed in Section \ref{sec:bg:existing_work}). The system then retrieves a free storage location from the recycling queue $Free\_Q$. If $Free\_Q$ is empty, a new location is allocated at the end of the file by default. Similar to the deletion phase, our method utilizes an efficient asynchronous I/O interface to write the relevant data of vertex $p$, including its vector $x_p$ and neighbor set $N_{out}(p)$, to the allocated position.

\begin{figure}[tbp]
    \centering
    \includegraphics[width=0.8\linewidth]{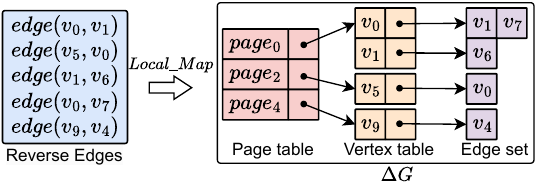}
    \vspace{-0.15in}
    \caption{An example of storing reverse edges in $\Delta G$.}
    \label{fig:deltaG}
    \vspace{-0.15in}
\end{figure}

\stitle{Page-Aware Cache Structure $\Delta G$}.
Similar to existing methods, our method adds reverse edges $\{edge(p', p) \mid p' \in N_{out}(p)\}$ for each inserted vertex $p$ to maintain good index accuracy. 
To avoid excessive random I/O overhead caused by immediately adding reverse edges upon insertion, we cache these reverse edges in a page-aware cache structure, $\Delta G$, as illustrated in Figure \ref{fig:deltaG}. 
By organizing reverse edges in pages, $\Delta G$ minimizes redundant updates to the same page. 
For each reverse edge, we first resolve the source vertex’s page ID in the disk index file using $Local\_Map$ (\eg resolving $edge(v_0, v_1)$ to $page_0$). 
It then looks up the corresponding vertex table in $\Delta G$’s page table (creating one if absent). Within the vertex table, it searches for the edge set corresponding to the source vertex ID (creating one if absent) and finally adds the target vertex ID to this set.

\etitle{Patch}.
To merge the updates in $\Delta G$ into the query index file, our method first identifies the pages that need to be updated based on $\Delta G$’s page table (e.g., $page_0$ in Figure \ref{fig:deltaG}). 
After loading the page (\ie $page_0$) from the query index file, the system sequentially processes the vertex table associated with $page_0$ in $\Delta G$. 
Taking vertex $v_0$ as an example, the system retrieves its pending reverse edges $\{v_1, v_7\}$ and merges them into the original neighbor set $N_{out}(v_0)$ in $page_0$, producing the updated set $N'_{out}(v_0)$. If $|N'_{out}(v_0)| \geq R$, the system applies a pruning algorithm (as described in Section \ref{sec:cmp}) to maintain the neighbor count within the limit $R$. After processing $v_0$, the system iterates through the subsequent vertices in the vertex table (e.g., $v_1$). Finally, once all vertex updates in $page_0$ are completed, the updated page is written back to the query index file via an asynchronous interface.

\subsection{I/O Overhead Analysis}
We compared the I/O overhead during batch updates between our method and the existing methods introduced in Section \ref{sec:bg:existing_work}.
For index updates, as described in Section \ref{sec:bg:existing_work}, \eat{the }existing methods perform two full index scans during the deletion and patch phases, leading to a total cost of $O(2(|X| + |G|))$. 
In contrast, our method identifies the affected vertices by scanning the separately stored lightweight topology file, thus reducing the cost to $O(|G|)$. Additionally, our method employs a page-level localized update mechanism, which only requires random I/O operations on the affected vertex set $V_{affect}$ and the incremental graph $\Delta G$, with a cost of $O(|V_{affect}| + |\Delta G|)$. Therefore, the total cost is approximately $O(|G| + |V_{affect}| + |\Delta G|)$.
In small-batch update scenarios where $|V_{affect}| \ll |X|$ and $|\Delta G| \ll |G|$, this design significantly reduces unnecessary I/O operations. 
In the insertion phase, both systems perform similar graph search-related random I/O operations. Overall, our method reduces a significant amount of I/O overhead in small-batch updates through decoupled lightweight topology and page-level localized update mechanisms. The experimental results in Figure \ref{fig:update_io} of Section~\ref{sec:expr:update_performance} validate the I/O efficiency of our method.

%% file: 4-2-compute.tex
\section{Lightweight Incremental Graph Repair}
\label{sec:cmp}

As described in Section \ref{sec:bg:existing_work}, existing systems handle updates by adopting neighbor repair algorithms to reduce computational overhead and avoid full graph reconstruction.
However, they still frequently trigger costly neighbor pruning, which significantly hinders update performance. In this section, we introduce two designs that reduce the neighbor pruning overhead caused by deletion and insertion operations.

\begin{figure}[tbp]
    \centering
    \begin{minipage}{0.5\textwidth}
        \begin{center}
        \subfloat[Delete]{\includegraphics[width = 0.45\linewidth]{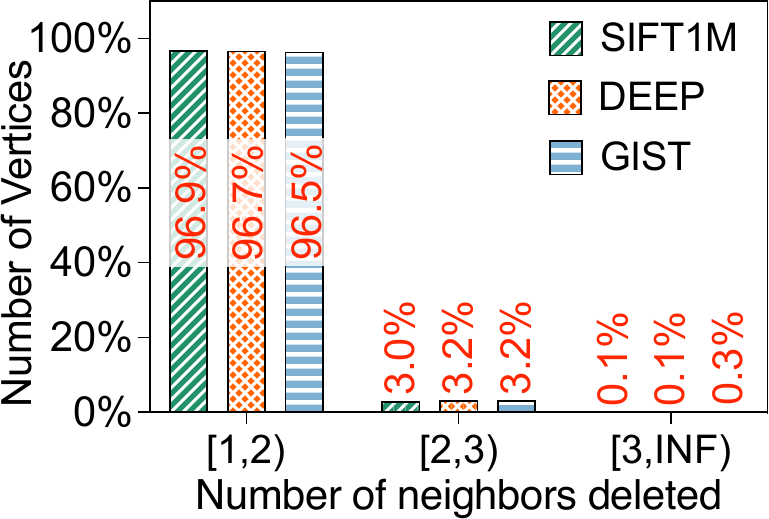}\label{fig:delete_nbr}}
        \hspace{0.1in}
        \subfloat[Patch]{\includegraphics[width = 0.45\linewidth]{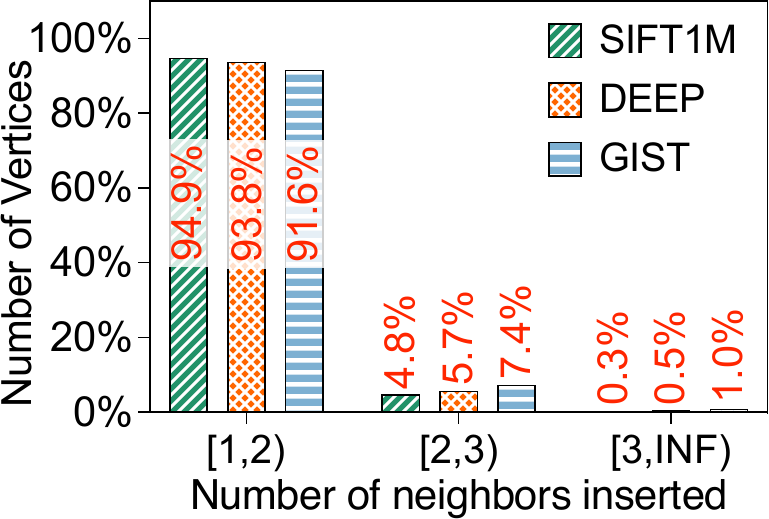}\label{fig:insert_nbr}}
        \captionsetup{font=small}
        \vspace{-0.15in}
        \caption{Distribution of the number of deleted and inserted neighbors.}
        \label{fig:prune_nbr_num}
        \end{center}
    \end{minipage}
        \vspace{-0.15in}
\end{figure}

\begin{figure*}[tbp]
    \centering
    \includegraphics[width=0.9\linewidth]{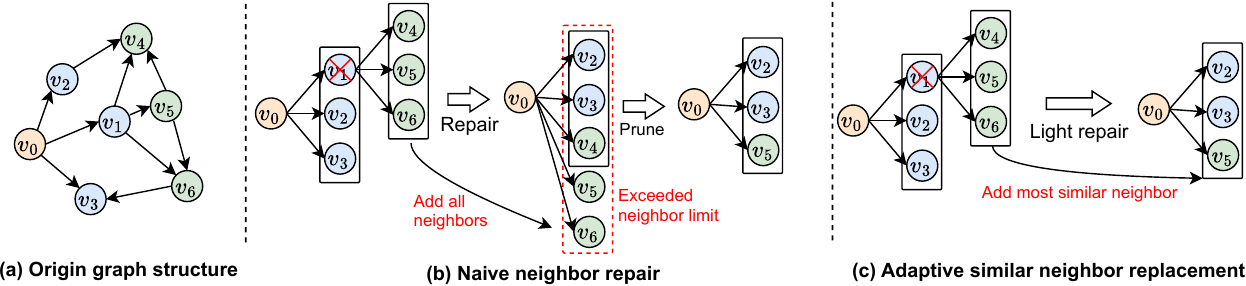}
    \vspace{-0.05in}
    \caption{Illustrate the deleted neighbor repair process for naive neighbor repair and our method respectively. (a) represents the original graph, while (b) and (c) depict the process of repairing the neighbors of $v_0$ after $v_1$ is deleted. The black rectangle indicates the neighbor limit $R=3$.
    }
    \label{fig:delete_adj}
    \vspace{-0.15in}
\end{figure*}

\subsection{Adaptive Similar Neighbor Replacement}
\label{sec:cmp:delete}

When a vertex's neighbors are deleted, existing methods repair the affected vertex to maintain the graph's navigability by connecting it to all neighbors of the deleted vertices, thereby compensating for the lost connectivity. 
This repair method enables the incremental maintenance of the graph structure within the index, avoiding the need to rebuild the index. However, this naive repair approach often results in an excessive number of neighbors for the vertices, which in turn triggers costly neighbor pruning.

\example{
Figure \ref{fig:delete_adj}(a) illustrates a simple example of an original graph-based index structure, where the edge lengths represent the distances between vectors, and the maximum number of neighbors per vertex is limited to $R = 3$. 
Figure \ref{fig:delete_adj}(b) demonstrates the process of the naive neighbor repair approach (\eg Algorithm \ref{alg:delete}) in repairing the neighbor set of vertex $v_0$ after the deletion of its neighbor $v_1$\eat{ is deleted}.  
First, the outgoing neighbors of $v_1$, $N_\text{out}(v_1) = \{v_4, v_5, v_6\}$, are added to the remaining neighbors of $v_0$ as candidate neighbors. Since the size of $v_0$'s candidate set expands to $5$, exceeding the limit $R$, an expensive pruning process is triggered. Ultimately, $v_0$ gets the pruned result set $\{v_2, v_3, v_5\}$ as its new neighbors.
}

\stitle{Observation}.
Based on the experiments described in Section \ref{sec:intro}, we analyze the number of neighbors deleted for each affected vertex \eat{in the}using real-world datasets. The results, as shown in Figure \ref{fig:delete_nbr}, reveal that \red{96\%} of the vertices experience the deletion of only one neighbor.  
However, as illustrated in Figure \ref{fig:delete_adj}b, even when only one neighbor is deleted, the naive neighbor repair approach (\eg Algorithm \ref{alg:delete}) adds all neighbors of the deleted neighbor to the candidate set. This often results in the size of the candidate set exceeding the neighbor limit, triggering the expensive pruning process every time.

Based on the observations above, we design an adaptive similar neighbor replacement algorithm \delalg, which performs lighter repairs to affected vertices, thereby avoiding the expensive neighbor pruning operations.

\stitle{Intuition}.
When an affected vertex has only a few neighbors deleted, it can still maintain good connectivity with other vertices in the graph through its remaining neighbors. 
A lightweight incremental algorithm should aim to reuse the existing neighborhood information as much as possible, compensating only for the connectivity lost due to the deletions. 
An intuitive and efficient approach is to replace the deleted vertex with similar \eat{vertices}ones from its neighbors, as similar neighbors often share similar connectivity information in graph-based indexes.
Therefore, the original neighborhood information of the deleted vertex can be accessed directly or indirectly through these similar neighbors.

However, careful consideration should be given to the number of similar neighbors used to replace a deleted vertex.
Our primary goal is to avoid triggering costly neighbor pruning when adding neighbors. 
Therefore, the number of similar neighbors for a deleted vertex should not exceed the available slots, i.e., the remaining neighbor space.
Additionally, it is important to consider the impact of each deletion. 
Intuitively, if a vertex has many neighbors, the impact of deleting one is relatively small as it is likely to retrieve the lost information through other neighbors. In such cases, the lost information can be compensated with fewer similar neighbors.
Conversely, if a vertex has a low original degree, the impact of deleting a neighbor is more significant, requiring more similar neighbors to replace it. 
Therefore, we design an adaptive similar neighbor replacement method to perform lightweight repairs on the neighbors of affected vertices, and it adaptively determines the number of neighbors to be replaced based on the number of neighbors of the vertex.

\begin{algorithm}
\small
\caption{\delalg($L_D, R, \alpha$)}
\label{alg:del_alg}
\KwIn{Graph $G(V, E)$ with $|V| = n$, set of vertices to be deleted $L_D$}
\KwOut{Graph on nodes $V'$ where $V' = V \setminus L_D$}

\Begin{
    \ForEach{$p \in V_{affect}$}{
        $\mathcal{D} \gets N_\text{out}(p) \cap L_D$\; \label{line:getd}
        $\mathcal{C} \gets N_\text{out}(p) \setminus \mathcal{D}$ \tcp*[r]{initialize candidate list} \label{line:getc}

        \If{$|\mathcal{D}| < T$}{
            $slot \gets R-|C|$ \tcp*[r]{available neighbor slots} \label{line:get-slot}
            $k_{slot} \gets max(\left\lfloor \left(\frac{slot}{|N_{\text{out}}(p)|}\right) \right\rfloor,1)$\; \label{line:get-kslot}
            \ForEach{$v \in \mathcal{D}$}{ \label{line:for}
                $\mathcal{C} \gets \mathcal{C} \cup SelectNearestNeighor(N_\text{out}(v)\setminus \mathcal{D}, k_{slot})$\;
            }
            $N_\text{out}(p) \gets \mathcal{C}$\; \label{line:get-nout}
        }
        \Else{
            \ForEach{$v \in \mathcal{D}$}{
                $\mathcal{C} \gets \mathcal{C} \cup (N_\text{out}(v) \setminus \mathcal{D})$\;
            }
    
            \If{$|\mathcal{C}| > R$}{
                $N_\text{out}(p) \gets \text{RobustPrune}(p, \mathcal{C}, \alpha, R)$\;
            }
            \Else{
                $N_\text{out}(p) \gets \mathcal{C}$\;
            }
        }
    }
}
\end{algorithm}

\stitle{Adaptive Similar Neighbor Replacement.}
Based on the above intuition, we design an adaptive similar neighbor replacement method, \delalg, to achieve a more lightweight repair of the affected vertices by avoiding the triggering of expensive neighbor pruning.
The core idea of \delalg is to replace deleted vertices with a small number of the most similar neighbors, rather than adding all neighbors. This approach avoids exceeding the neighbor size limit and prevents the expensive distance computations triggered by pruning. The pseudocode of \delalg is shown in Algorithm \ref{alg:del_alg}.  

Specifically, for each affected vertex $p$, \delalg first retrieves its deleted neighbors set $\mathcal{D}$ and the non-deleted neighbors set $\mathcal{C}$. 
If the size of $\mathcal{D}$ is smaller than a user-defined threshold $T$, the following steps are performed: 
First, we count the number of available neighbor slots in the neighbor space of vertex $p$ (denoted as $slot$).
Next, we determine the number of similar neighbors needed to replace each deleted vertex (denoted as $K_{slot}$)\eat{ is determined}.
The algorithm calculates $k_{slot}$ based on the current number of available slots ($slot$) and the original neighbors of $p$, using the formula: $k_{slot} \gets \max\left(\left\lfloor \frac{slot}{|N_{\text{out}}(p)|} \right\rfloor, 1 \right)$.
Then, for each deleted neighbor $v \in \mathcal{D}$, we select the $k_{slot}$ most similar (\ie shortest edge length) undeleted vertices from $N_{out}(v)$ and add them to $\mathcal{C}$.
Since the total number of deleted neighbors is always less than or equal to the original number of neighbors, i.e., $|\mathcal{D}| \leq |N_{\text{out}}(p)|$, the number of deleted neighbors after replacement satisfies $k_{slot} \times |\mathcal{D}| \leq slot$, thus ensuring $|\mathcal{C}| \leq R$ and preventing the need for neighbor pruning.
After all \eat{the }vertices in $\mathcal{D}$ have been processed, $\mathcal{C}$ \eat{is used as}becomes the new $N_{out}(p)$ for vertex $p$. If the size of $\mathcal{D}$ is greater than or equal to $T$, we follow the same process as described in Algorithm \ref{alg:delete}\eat{ is followed}.

\example{
Figure \ref{fig:delete_adj}c illustrates the process of \delalg repairing the neighbors of $v_0$ after $v_1$ is deleted, as shown in Figure \ref{fig:delete_adj}a. First, \delalg collects the remaining neighbor set of $v_0$, denoted as $\mathcal{C}=\{v_2,v_3\}$. It then examines $N_{out}(v_1)$ and selects \eat{a vertex $v_5$}{$v_5$, a vertex} that has not been deleted and is nearest to $v_1$. 
Finally, $v_5$ is added to $\mathcal{C}$, forming the new neighbor set $N_{out}(v_0)$. Since $|N_{out}(v_0)| \leq R$, this approach avoids triggering costly neighbor pruning \eat{compared to}{in contrast to} Figure \ref{fig:delete_adj}b.
}

As introduced in Section \ref{sec:io:update}, the patch phase inserts the cached reverse edges from $\Delta G$ into the corresponding vertex neighbor sets. 
However, \eat{added}{the addition of} reverse edges often cause the number of neighbors of updated vertices to exceed the limit $R$, thereby triggering costly pruning.
Similar to Figure \ref{fig:delete_nbr}, we count the number of reverse edges added when vertices trigger pruning in three real-world datasets.
As shown in Figure \ref{fig:insert_nbr}, the results indicate that \red{90\%} of the pruning is triggered by the addition of only a single edge. 
The existing batch update method (such as \freshdiskann) enforces a strict neighbor limit, that is, when a vertex's neighbor set is full, adding even a single edge can cause the set to exceed the limit $R$, triggering an expensive pruning process.

\stitle{Relaxed Neighbor Limit.} 
To reduce the significant computational cost caused by adding a small number of reverse edges, we implement a relaxed neighbor limit during this addition process. Specifically, each vertex's neighbors are constrained by two parameters: a strict neighbor limit $R$ and a relaxed neighbor limit $R'$, where $R \leq R'$. When storing the neighbors of each vertex on disk, $R'$ slots are allocated, with $R' - R$ reserved as additional space.  
During pruning, the strict neighbor limit is enforced to ensure that the number of neighbors does not exceed $R$.
When adding reverse edges to a vertex, the relaxed neighbor limit is applied, allowing up to $R'$ neighbors before pruning is triggered. 

Theoretically, the larger $R' - R$ is, the fewer pruning operations are triggered. However, it is important to note that if $R' - R$ is too large, the number of neighbors will increase, \eat{which may result in}{potentially leading to} higher computational overhead during search processing due to the need to traverse more neighbors.  
To balance these trade-offs, we set $R'$ based on the results in Figure \ref{fig:insert_nbr}—where \red{90\%} of pruning is triggered by the addition of a single edge—and defaults $R'$ to $R + 1$. This setting significantly reduces the number of pruning operations while minimally impacting search efficiency.

Moreover, adopting a relaxed neighbor limit increases the neighbor storage space by $R' - R$, potentially affecting the file size. However, since $R' - R$ is typically small, the reserved space constitutes only a negligible portion of the total index size. Furthermore, the disk-based index storage is page/block-aligned, and there is usually some unused space in each page/block \cite{DiskANN,FreshDiskANN,SecondTierMemory}. Our experiments in Section \ref{sec:expr:index_size} reveal that the space used for $R' - R$ neighbors mostly comes from this unused space, resulting in no noticeable increase in file size.

\subsection{Computation Overhead Analysis}
We briefly analyze the computational overhead of our method compared to the existing methods \eat{introduced}described in Section \ref{sec:bg:existing_work}.
During the deletion phase, \eat{the }existing methods use Algorithm \ref{alg:delete} to repair a vertex when its neighbors include deleted vertices, triggering pruning \eat{each time, }with a complexity of $O(|\mathcal{C}|^{2} \cdot d)$, where $|\mathcal{C}| \approx |\mathcal{D}| \cdot R + R$, and $d$ is the vector dimension. 
In contrast, for a vertex with only a few deleted neighbors (as shown in Figure \ref{fig:prune_nbr_num}, accounting for 90\% of cases), we use a similar neighbor replacement method (i.e., Algorithm \ref{alg:del_alg}), ensuring that the number of neighbors after repair does not exceed $R$, thus avoiding \eat{triggering }pruning. 
The complexity of this method arises from finding similar neighbors, i.e., calculating the distance between each deleted vertex and its neighbors,
which has a complexity of $O(|\mathcal{D}| \cdot R \cdot d)$. 
By avoiding neighbor pruning, we significantly reduce computational overhead.
For the patch phase, we allow the number of neighbors to exceed $R$ but remain less than or equal to $R'$ when adding reverse edges, thus reducing the frequency of pruning triggers compared to the existing methods. 
Overall, our method reduces the number of neighbors pruning in both phases and the experimental results in Figure \ref{fig:update_cmp} of Section \ref{sec:expr:update_performance} also validate that our method has a lower probability of triggering pruning.

%% file: 4-3-impl.tex
\section{Implementation}
\label{sec:impl}

We implement an ANNS system prototype, \oursys, based on the topology-aware index on top of \freshdiskann \cite{FreshDiskANN}. 
\oursys also incorporates the characteristics of our strategy to design the following features that efficiently support the system's I/O operations and concurrent safety.
\stitle{Asynchronous I/O}.
For our fine-grained page-level localized
update method, this paper designs an efficient I/O controller based on the Linux asynchronous I/O library (libaio). The controller utilizes asynchronous, non-blocking, and parallel I/O mechanisms to \eat{enable}facilitate efficient concurrent access to specified index file pages. Its core process consists of three stages: request preprocessing, batch submission, and event polling. 
In the first stage, asynchronous I/O control blocks (iocbs) are constructed using the \textit{io\_prep\_pread} and \textit{io\_prep\_pwrite} interfaces, with explicit request parameters. 
In the second stage, the \textit{io\_submit} system call is used to submit the batch requests to the kernel I/O scheduling queue in a non-blocking manner, significantly reducing the context switch overhead between user space and kernel space. This allows a single thread to handle multiple I/O operations concurrently, thus improving overall throughput. 
In the final stage, the \textit{io\_getevents} interface actively polls the completion event queue, batches the collected completed I/O results, and updates the corresponding index pages using the localized update strategy.
This design not only optimizes I/O performance but also provides robust support for fine-grained updates.

\stitle{Page-Level Concurrency Control}.
\oursys supports concurrent vector search and vector updates, making it essential to ensure the safety of these operations in a concurrent read-write environment. In the fine-grained update mode, each update operation typically involves small file pages, and the update operations for individual pages are executed quickly. Based on this, we design a page-level fine-grained concurrency control mechanism with page-level read-write locks to ensure the safety of system reads and writes. It manages concurrent access at the page level, ensuring that when multiple queries and update operations are performed simultaneously, data contention and inconsistency issues are effectively prevented.

%% file: 5-eval.tex
\section{Evaluation}
\label{sec:expr}

\begin{table}[t!]
\centering
\small
\caption{Dataset description. $T$ and $D$ denote the dimensionality and data type of vectors. 
} 
\label{tab:dataset}
\begin{tabular}{ccccccc}
\hline
\textbf{Dataset} & \textbf{$T$} & \textbf{$D$} & \textbf{\# Vector} & \textbf{\# Query} & \textbf{Contents} \\ \hline
SIFT1M              & float  & 128   & 1,000,000        & 10,000  & Image              \\ 
Text2Img            & float  & 200   & 1,000,000        & 1,000   & Image \& Text      \\ 
DEEP                & float  & 256   & 1,000,000        & 1,000   & Image              \\ 
Word2Vec            & float  & 300   & 1,000,000        & 1,000   & Word Vectors        \\ 
MSONG               & float  & 420   & 994,185          & 1,000   & Audio               \\ 
GIST                & float  & 960   & 1,000,000        & 1,000   & Image               \\ 
MSMARC              & float  & 1024  & 1,000,000        & 1,000   & Text                \\
SIFT1B              & uint8  & 128   & 1,000,000,000    & 1,000   & Image               \\ \hline 
\end{tabular}
\end{table}

\subsection{Evaluation Setup.}

\stitle{Evaluation Platform.}
All experiments are conducted on a server equipped with an Intel(R) Xeon(R) Gold 6248R CPU @ 3.00GHz, 320GB DDR4 memory (@3200MT/s), 48 cores and 2 SSDs of 1.7TB, capable of achieving up to 500MBps sequential read and write performance. The operating system is Ubuntu 22.04 LTS, and the compiler used is GCC 11.4.0.

\stitle{Datasets.}
We evaluate our system using eight public real-world datasets, with detailed statistics provided in Table \ref{tab:dataset},
including SIFT1M \cite{sift-link}, Text2Img \cite{text2img-link}, DEEP \cite{deep-link}, Word2Vec \cite{deep-link}, MSONG \cite{deep-link}, GIST \cite{sift-link}, MSMARC \cite{deep-link} and SIFT1B \cite{sift-link}.
They cover diverse dimensions, sizes, and content types, and have been widely adopted in prior studies
\cite{DiskANN,HNSW,spfresh,Milvus,NSG,FreshDiskANN} for benchmark ANNS systems.

\stitle{Compared Systems and Parameters.}
In the experiments, we compare the following three systems:
\begin{itemize}[leftmargin=*]
\item[$\bullet$]
\textbf{\freshdiskann} \cite{FreshDiskANN} is the state-of-the-art graph-based streaming ANNS system on disk, developed and open-sourced by Microsoft. It adopts an out-of-place update strategy by incrementally repairing the graph structure in the index and rebuilding the entire index file, as introduced in section \ref{sec:bg:existing_work}. 
Its settings are as follows,
the candidate neighbor limit $MAX_C$ for graph construction is set to \red{500}, the neighbor limit $R$ for each vertex is set to \red{32}, 
the insertion priority queue length limit $L\_build$ is set to \red{75},
and the query priority queue length limit $L\_search$ is set to \red{120}.
By following most of the existing studies \cite{spfresh,FreshDiskANN}, we use 2 threads for search, 3 threads for insertion, 1 thread for deletion, and 10 threads for index batch update.

\item[$\bullet$]
\textbf{\ipdiskann} \cite{IP-DiskANN} is a recently proposed algorithm for dynamically updating DiskANN indexes. This algorithm avoids a full index scan by querying the $l_d$ closest vertices to the deleted vertex to identify its incoming neighbors. However, it does not guarantee that all incoming neighbors of the deleted vertex are found, leaving some dangling edges pointing to the deleted point. 
Therefore, it is necessary to remove these dangling edges through periodic additional full-graph scans.
Furthermore, to reduce the computational overhead of pruning incoming neighbors, it only selects $c$ neighbors of the deleted vertex to connect with its incoming neighbors. Unlike our algorithm, this approach may still exceed the neighbor limit and trigger pruning. Although \ipdiskann is an algorithm, we have reproduced its code under the localized update strategy of \oursys for performance comparison with different pruning algorithms. We also ensured that it reused all of our I/O and disk update optimizations. In addition to the parameters common to \freshdiskann, the extra parameter settings refer to the configurations in its paper, such as $l_d = 128$ and $c = 3$.

\item[$\bullet$] 
\textbf{\oursys } is our system, which reduces the large amount of unnecessary I/O overhead through fine-grained page-level localized updates and employs a similar neighbor replacement method to minimize computational costs. 
It is developed based on \freshdiskann, maintaining the same parameter settings, such as $MAX_C$, $R$, $L\_build$ and $L\_search$. Additionally, for \oursys, the newly introduced deletion threshold $T$ and relaxed neighbor limit $R^{'}$ are set by default to $T = \red{2}$ and $R^{'} = \red{33}$, unless otherwise specified.

\end{itemize}

\stitle{Metrics.}
\oursys is designed to achieve fast updates of the index structure, so it primarily focuses on update performance. For update performance, it is measured by update throughput, indicating the number of insertions/deletions the system can handle per second. Additionally, to validate the impact of Greater on index quality and other overheads, we also test several other metrics. Search accuracy is evaluated using k-recall@k, with \(k\) set to 10. Search performance is measured using tail latency metrics (such as P90, P95, P99, and P99.9). I/O overhead tracks the amount of disk I/O generated by read and write operations, while space overhead represents the disk space occupied by the index.

\begin{figure*}[tbp]
    \centering
    \includegraphics[width=0.9\linewidth]{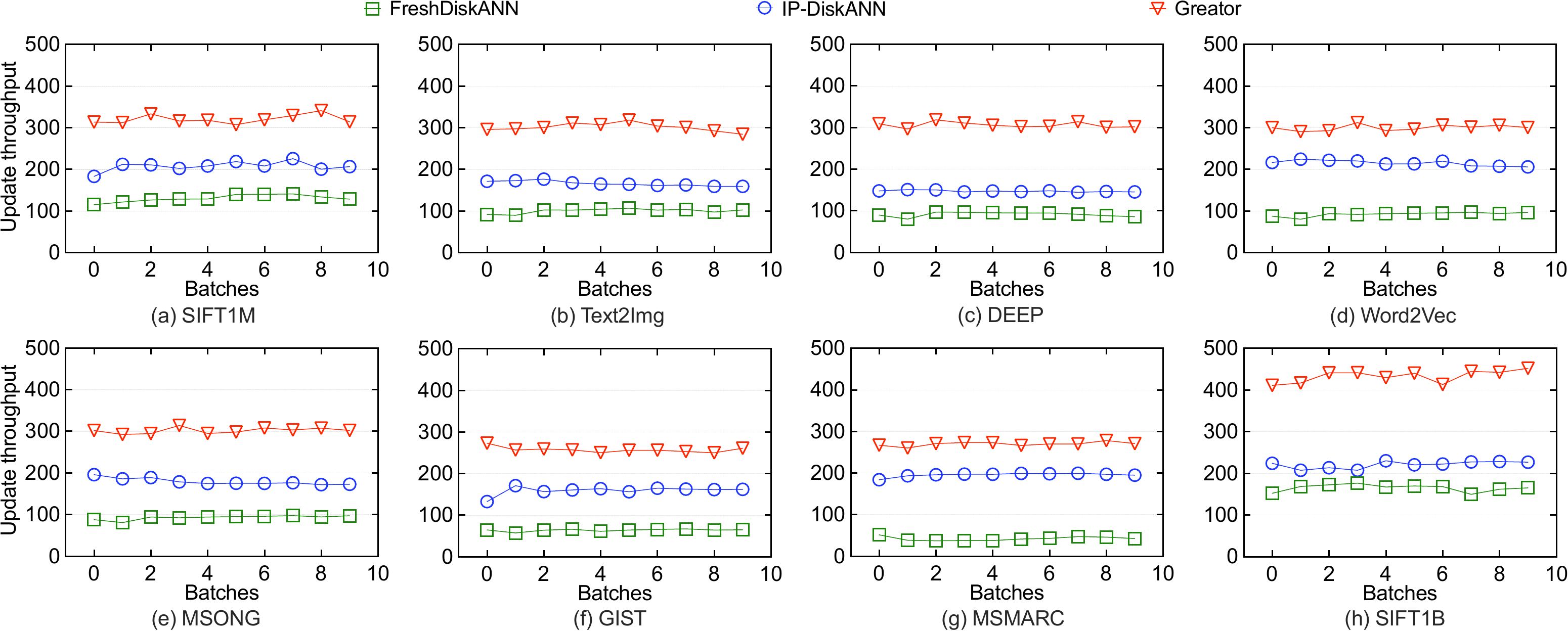}
    \captionsetup{font=small}
    \vspace{-0.15in}
    \caption{Update throughput comparison. 
    }
    \label{fig:update_ops}
\end{figure*}

\subsection{Update Performance}
\label{sec:expr:update_performance}
We first evaluate the update performance of different systems on \red{eight} real-world vector datasets (details are shown in Table \ref{tab:dataset}). 
We adopt the workload design of \freshdiskann. Specifically, we first use \red{99\%} of each dataset to statically construct the base index. Then, we perform batch updates by deleting 0.1\% of the existing vectors and inserting 0.1\% of the vectors from the remaining \red{1\%} of the dataset in each batch.

\stitle{Update Throughput Comparison}.
To evaluate the update performance of different systems, we compare their update throughput over \red{10} consecutive batch updates,
as shown in Figure~\ref{fig:update_ops}. 
We can see that \oursys achieves higher throughput than \freshdiskann and \ipdiskann across all datasets. Specifically, \oursys improves update throughput by \red{2.47$\times$-6.45$\times$} (on average \red{3.55$\times$}) compared to \freshdiskann, 
\red{1.38$\times$-2.08$\times$} (on average \red{1.69$\times$}) compared to \ipdiskann.
Compared with \freshdiskann, the primary performance gains of \oursys come from our topology-aware localized update strategy, such as reducing unnecessary I/O overhead through a lightweight topology and a localized update method, and reducing computational overhead with a similar neighbor replacement method. 
Compared to \ipdiskann, our gains mainly come from the approximate neighbor replacement method, which effectively avoids pruning triggered by the number of neighbors exceeding the threshold ($R$). The primary goal of \ipdiskann is to reduce the number of neighbors involved in pruning, but it still triggers a significant amount of pruning.

\begin{figure}
    \centering
    \includegraphics[width=0.9\linewidth]{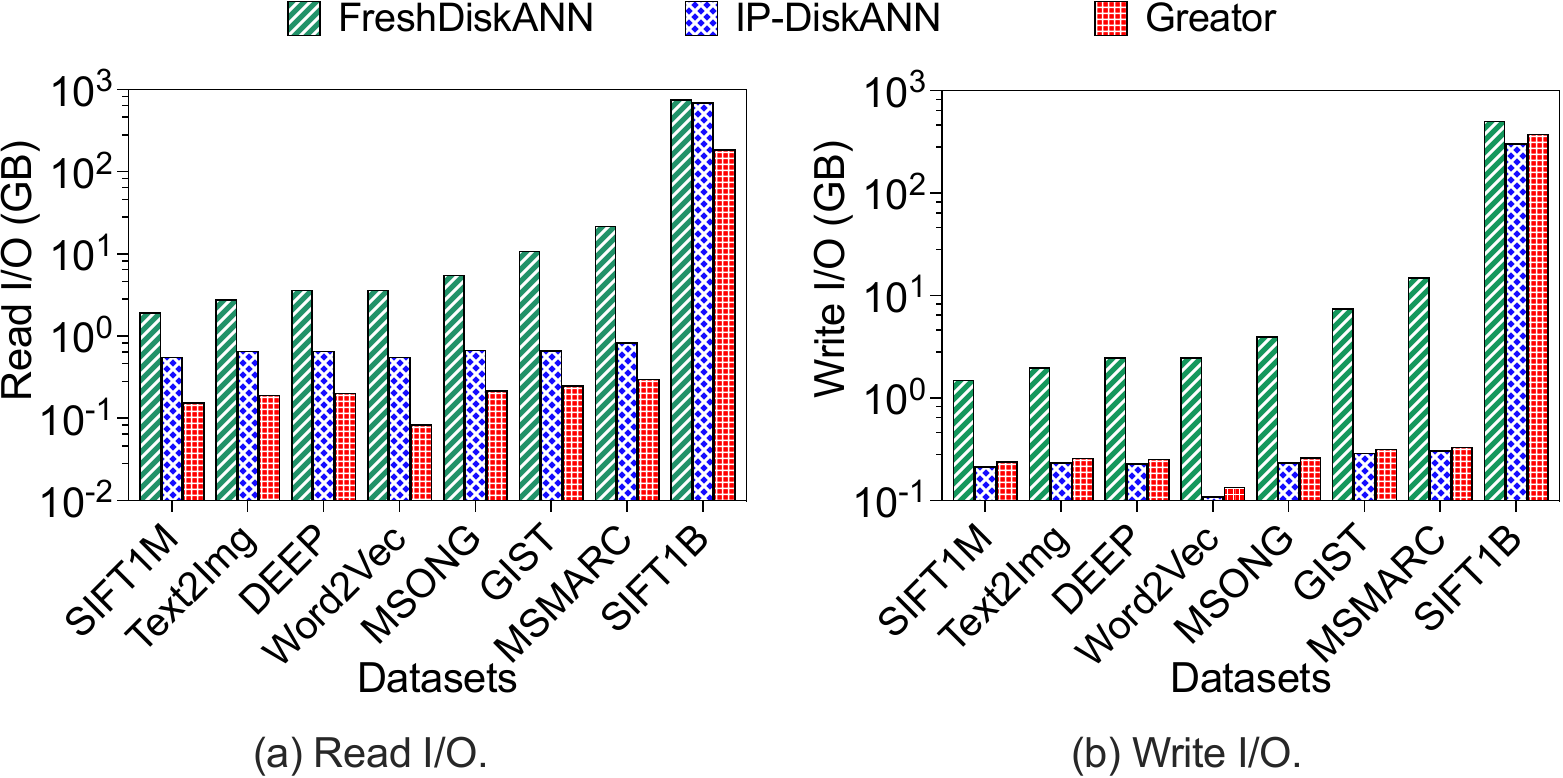}
    \vspace{-0.15in}
    \caption{I/O amount comparison.}
    \label{fig:update_io}
    \vspace{-0.15in}
\end{figure}

\stitle{I/O Amount Comparison}.
We also measure the read and write I/O amount generated by different systems, and the results are shown in Figure \ref{fig:update_io}. \oursys achieves a lower I/O amount compared to the other two competitors. Specifically, compared to \freshdiskann, \oursys reduces read I/O amount by \red{4.06$\times$-73.07$\times$} (average \red{29.28$\times$}), and write I/O amount by \red{1.34$\times$-44.88$\times$} (average \red{15.85$\times$}). 
The main benefit comes from \oursys, which avoids scanning the entire graph to identify affected vertices by using a lightweight topology and reduces unnecessary read and write I/O amount with page-level localized update methods. Compared to \ipdiskann, \oursys reduces read I/O amount by \red{2.65$\times$-6.56$\times$} (average \red{3.65$\times$}), while write I/O remains nearly the same. In terms of reading, \ipdiskann uses an approximate search for each deleted vertex to find possible incoming neighbors in the index, avoiding the overhead of scanning the index file. However, this ANNS search process still results in significant I/O. 
In terms of writing, \ipdiskann is reproduced on \oursys and leverages the localized update strategy of \oursys to achieve similar write performance, which further demonstrates the I/O effectiveness of our localized update approach.

\begin{figure}
    \centering
    \includegraphics[width=0.9\linewidth]{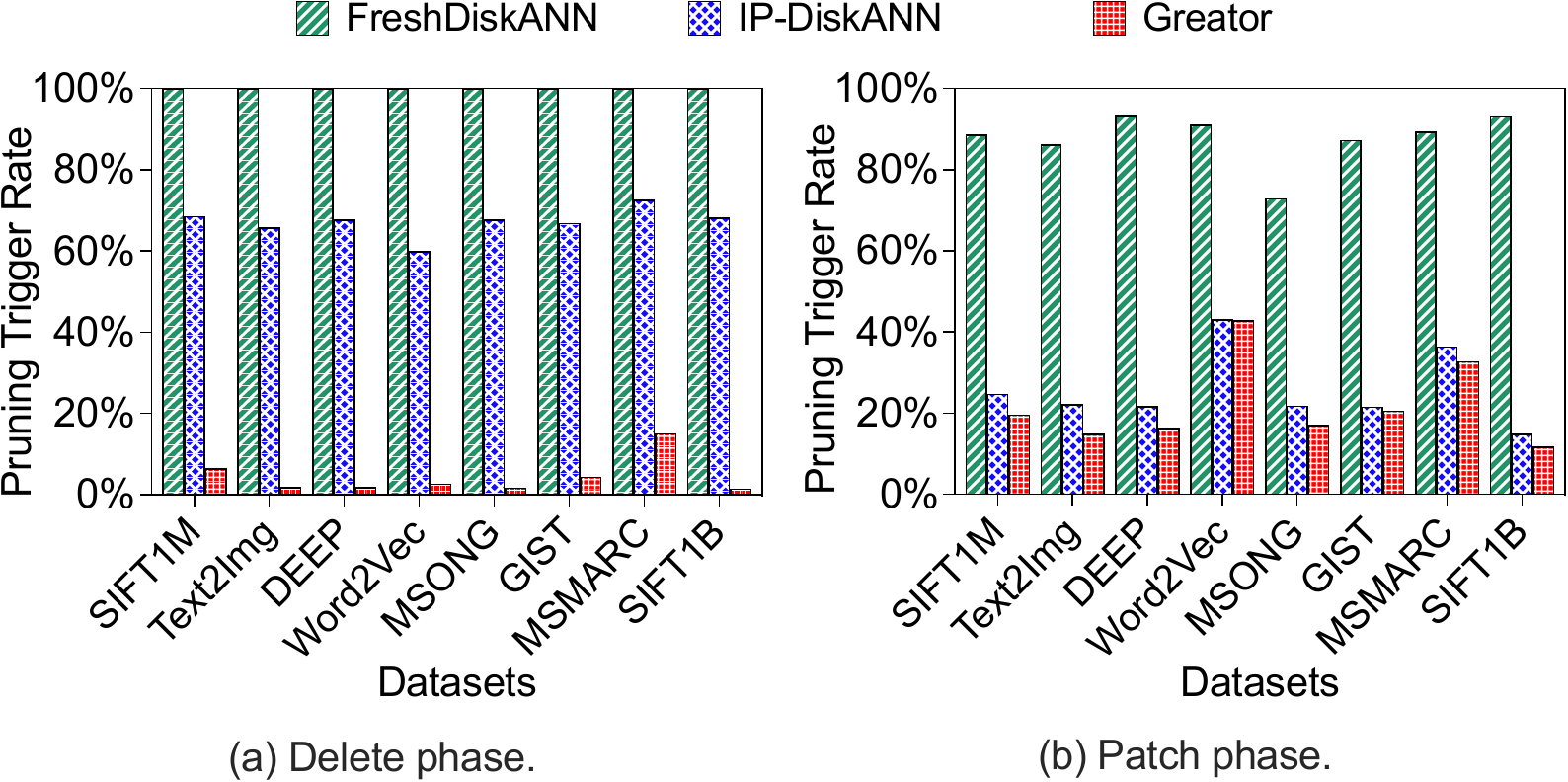}
    \vspace{-0.15in}
    \caption{Pruning count comparison.}
    \label{fig:update_cmp}
    \vspace{-0.15in}
\end{figure}

\stitle{Pruning Count Comparison}.
We further measured the proportion of pruning operations triggered during the delete and patch phases for different systems, and the results are shown in Figure \ref{fig:update_cmp}. As can be seen, both \oursys and \ipdiskann trigger significantly fewer pruning operations across all datasets. 
Specifically, during the deletion phase, the neighbor pruning rate in \oursys is on average reduced by \red{84.83\%-98.44\%} (average \red{95.51\%}) compared to \freshdiskann, and reduced by \red{57.20\%-66.67\%} (average \red{62.73\%}) compared to \ipdiskann. This is mainly due to \oursys's adoption of a similar neighbor replacement method, which significantly avoids triggering pruning. \ipdiskann reduces the number of triggers compared to \freshdiskann by reconstructing only a subset of neighbors during the repair process, thereby avoiding some unnecessary pruning operations. 
During the patch phase, \oursys reduces neighbor pruning by \red{48.30\%-81.57\%} (average \red{65.87\%}) compared to \freshdiskann, while \ipdiskann behaves similarly to \oursys. Both \oursys and \ipdiskann benefit from our relaxed neighbor constraints, resulting in a significant reduction in neighbor pruning compared to \freshdiskann.

\stitle{Update Stability}.
As shown in Figure~\ref{fig:update_ops}, \oursys, \ipdiskann, and \freshdiskann maintain stable update throughput across consecutive batch updates on different datasets. This demonstrates that these systems can effectively handle streaming updates and are suitable for real-world online applications.

\subsection{Index Quality}

Although multiple designs in \oursys, such as the localized update strategy and the similar neighbor replacement strategy, significantly improve update performance, it is still necessary to evaluate whether \oursys can maintain index quality comparable to \freshdiskann. Based on the index generated after each batch update in Section~\ref{sec:expr:update_performance}, we assess query performance using the query vector sets provided by each dataset (see Table~\ref{tab:dataset}). The following presents the experimental results in terms of search accuracy and search latency.
The following presents the experimental results on search accuracy and search latency.

\begin{figure*}[tbp]
    \centering
    \includegraphics[width=0.9\linewidth]{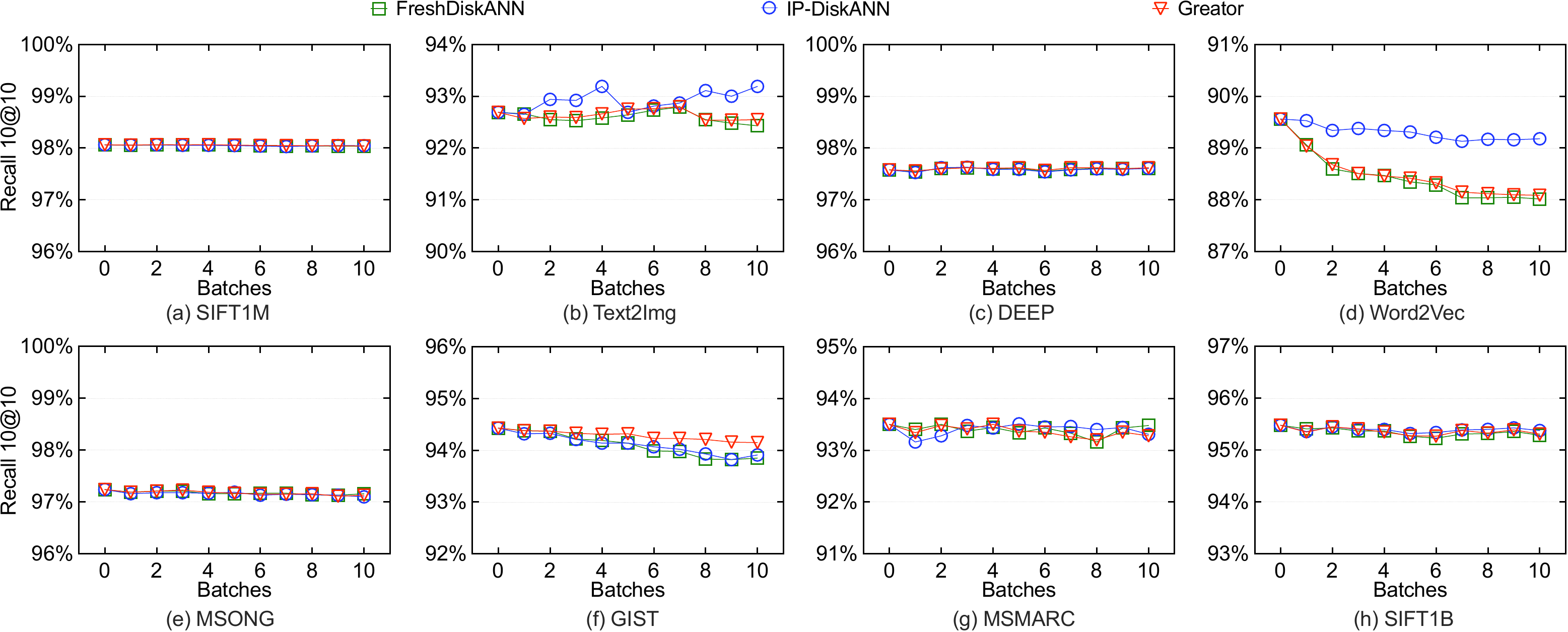}
    \captionsetup{font=small}
    \vspace{-0.15in}
    \caption{Recall comparison.
    }
    \label{fig:recall_stability}
\end{figure*}

\begin{figure*}[tbp]
    \centering
    \includegraphics[width=0.9\linewidth]{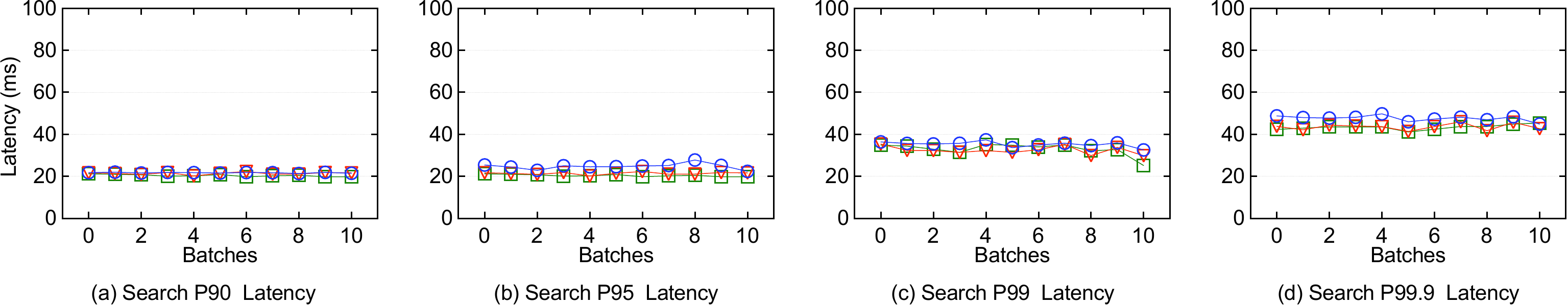}
    \captionsetup{font=small}
    \vspace{-0.15in}
    \caption{Search latency comparison on MSMARC.
    }
    \label{fig:search_latency}
\end{figure*}

\stitle{Search Accuracy Comparison}.
As shown in Figure~\ref{fig:recall_stability}, we present the query recall of different systems after consecutive updates across all datasets. 
Only on the Word2Vec dataset, \ipdiskann shows higher recall, which may be due to its connection of incoming neighbors between some deleted vertices. On the other datasets, all three systems perform similarly.
This result indicates that \oursys's similar neighbor replacement algorithm not only significantly improves update performance by reducing neighbor pruning but also maintains high search accuracy during updates.

\stitle{Search Latency Comparison}.
The similar neighbor replacement method and the relaxed neighbor limit strategy adopted by \oursys significantly improve update performance but also introduce structural differences compared to \freshdiskann. For example, the relaxed neighbor limit strategy may increase the number of neighbors per vertex, leading to more neighbor traversals during search and potentially increasing query latency. To evaluate the search performance of \oursys, we measure the tail query latency (including P90, P95, P99, and P99.9) after consecutive updates on the MSMARC dataset, with results shown in Figure~\ref{fig:search_latency}. 
The results indicate that the tail latency of \oursys remains nearly identical to that of \freshdiskann, demonstrating that the similar-neighbor replacement algorithm and the relaxed neighbor limit strategy enhance update performance while maintaining high navigation efficiency in the index.

\begin{figure}
    \centering
    \includegraphics[width=1\linewidth]{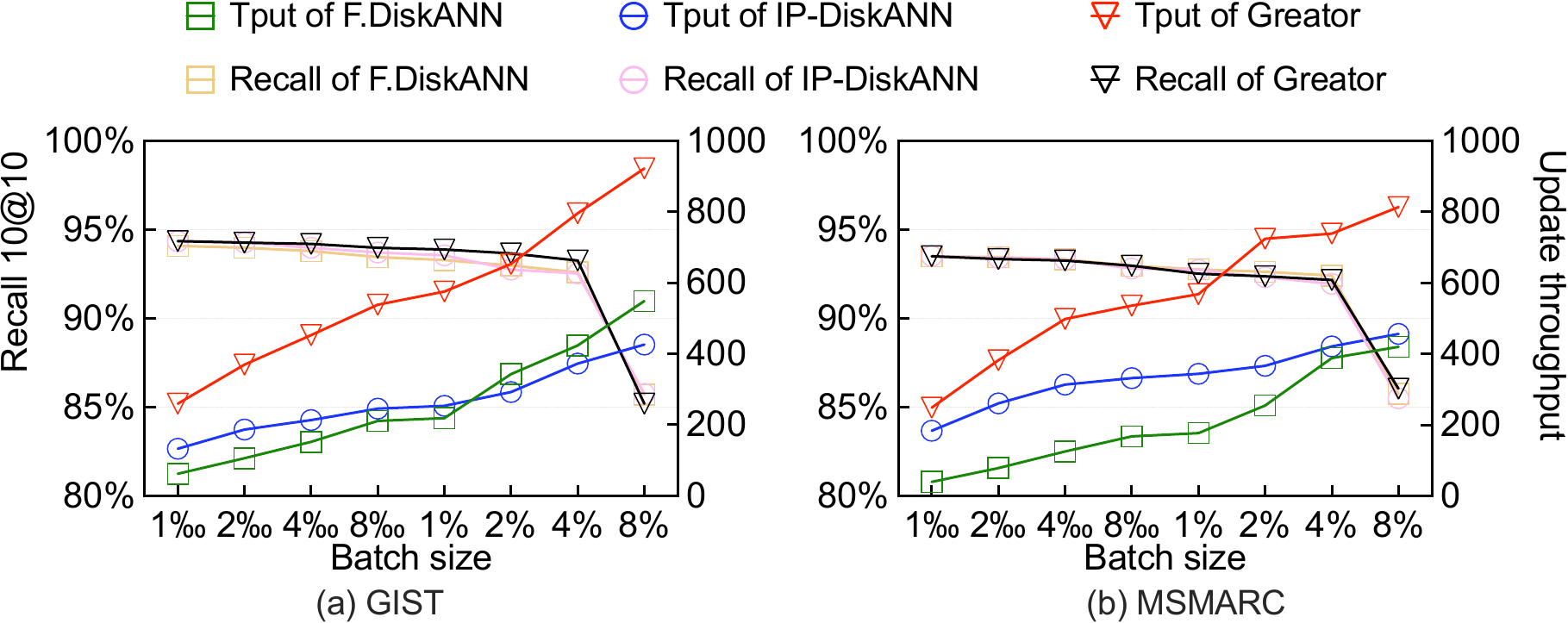}
    \vspace{-0.25in}
    \caption{Varying batch size of updates. 
    }
    \vspace{-0.15in}
    \label{fig:vary_batch_size}
\end{figure}

\subsection{Sensitivity to Batch Size}
\label{sec:expr:vary_batch_size}
We evaluate the update performance and search accuracy of different systems under different batch sizes on the GIST and MSMARC datasets, with results shown in Figure~\ref{fig:vary_batch_size}. As the batch size gradually increases from \red{0.1\%} to \red{8\%} of the dataset, \oursys, \ipdiskann and \freshdiskann exhibit a steady improvement in update performance, while \oursys consistently maintains a stable performance advantage.  
As discussed in Section~\ref{sec:intro}, the recall rate decreases as the batch size increases, with a particularly sharp decline when the batch size reaches \red{8\%}. 
Notably, there exists an intersection point between recall and update performance, and the intersection point of \oursys is significantly higher than that of \freshdiskann and \ipdiskann. This indicates that \oursys achieves a better trade-off between performance and accuracy than them.

\subsection{Performance Gain}
\label{sec:expr:gain}

\begin{figure}
    \centering
    \includegraphics[width=0.8\linewidth]{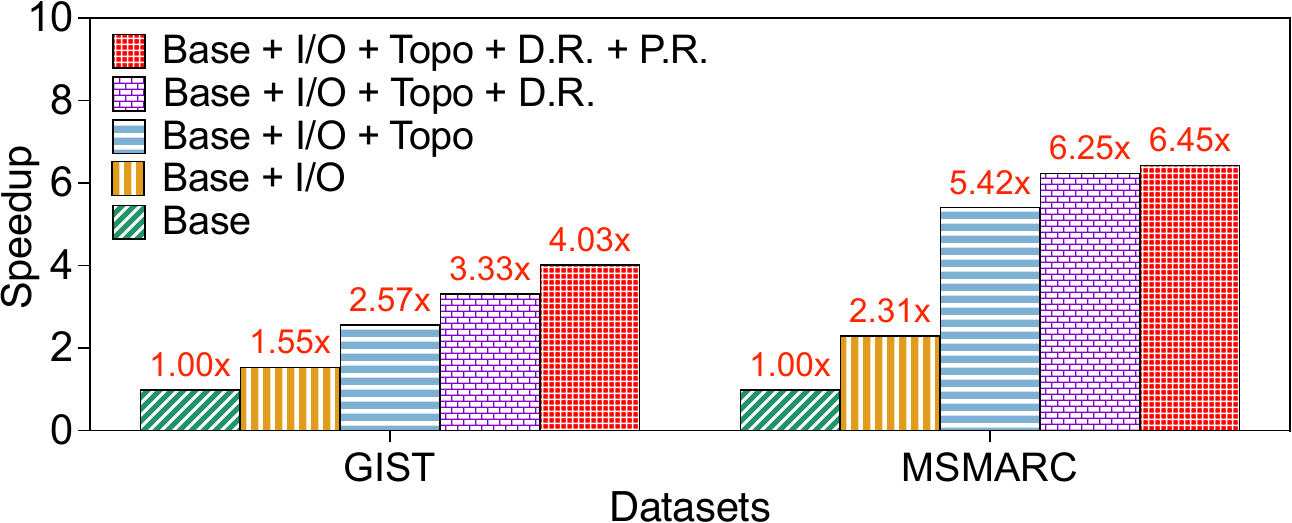}
    \vspace{-0.12in}
    \caption{Performance gain.
    }
    \label{fig:perform_gain}
    \vspace{-0.2in}
\end{figure}

To analyze the performance gains of each design in \oursys, we start from the baseline version (\freshdiskann) and incrementally integrate the novel designs of \oursys, evaluating them on two real-world datasets (GIST and MSMARC). Figure~\ref{fig:perform_gain} presents the normalized speedup.  
First, incorporating the localized update strategy (+I/O), which updates only the affected vertices, achieves a speedup of \red{1.55$\times$} and \red{2.31$\times$} on the two datasets, respectively. 
Next, introducing the lightweight topology (+Topo) improves performance to \red{2.57$\times$} and \red{5.42$\times$} by avoiding full index file scans. Subsequently, integrating the similar neighbor replacement method (+D.R.) further enhances performance to \red{3.33$\times$} and \red{6.25$\times$} by reducing neighbor pruning. Finally, applying the relaxed neighbor limit strategy (+P.R.) further decreases the computational overhead of pruning, ultimately achieving a performance boost to \red{4.03$\times$} and \red{6.45$\times$}.  
These results demonstrate that all core design choices in \oursys effectively contribute to improving update performance.

\subsection{Cost of Lightweight Index Topology}
\label{sec:expr:index_size}

Although the lightweight topology in \oursys significantly reduces I/O and improves update performance, it also introduces additional space and time overhead. 
Next, we conduct an experimental analysis of these two types of costs.

\stitle{Space Cost}.
We measure the storage space overheads of \oursys compared to \freshdiskann (\ipdiskann and \freshdiskann are consistent), as illustrated in Figure \ref{fig:index_size}. It can be observed that the index file size of \oursys is on average \red{1.15$\times$} that of \freshdiskann, primarily due to the additional storage required for the lightweight topology in \oursys.
Notably, these files are stored on inexpensive and high-capacity disks, making the slight increase in space overhead generally acceptable. 
Furthermore, as the dataset dimensionality increases (when the data types are the same), the difference in index file size between \oursys and \freshdiskann gradually decreases, resulting in a less pronounced space amplification effect.

\begin{figure} 
\centering
    \begin{minipage}{0.45\linewidth}
        \subfloat{\includegraphics[width = 1\linewidth]
            {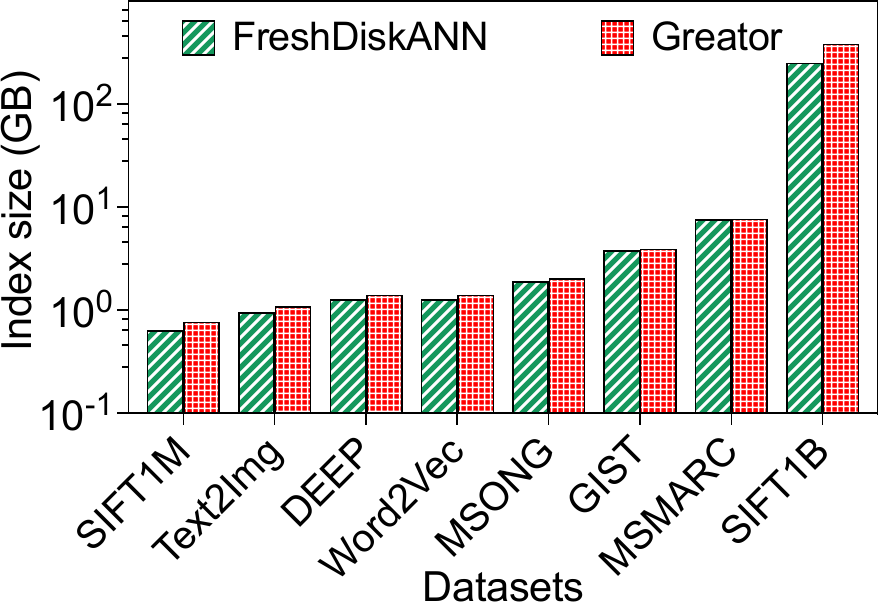}\label{fig:index_size} \hspace{0in}}
        \vspace{-0.12in}
        \captionsetup{font=small}
        \caption{Space cost comparison. 
        }
        \label{fig:index_size}
    \end{minipage}
    \ \ \
    \begin{minipage}{0.45\linewidth}
        \subfloat{\includegraphics[width = 1\linewidth]
            {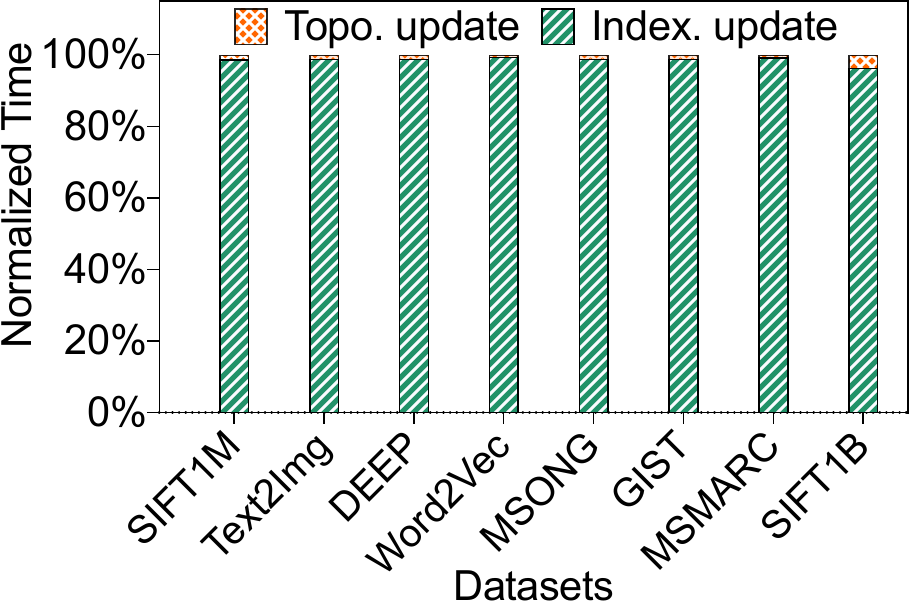}\label{fig:topo_update} \hspace{0in}}
        \vspace{-0.12in}
        \captionsetup{font=small}
        \caption{
        Time cost comparison. 
        }
        \label{fig:time_cost}
    \end{minipage}

  \vspace{-0.15in}
\end{figure}

\stitle{Time Cost}.
We measure the proportion of total update time spent maintaining the lightweight topology in \oursys, with results shown in Figure~\ref{fig:time_cost}. It can be observed that the maintenance overhead accounts for only \red{0.56\%-3.68\%} of the total update time, exerting a very minimal impact on the system's update performance.
Furthermore, as discussed in Section~\ref{sec:expr:gain}, the lightweight index topology plays a critical role in improving system performance, further validating the efficiency of this design.

%% file: 6-related.tex
\section{Related Work}
\label{sec:relatedwork}

\stitle{Vector Indexes.}
The widespread application of ANNS has led to significant research in vector indexing \cite{HNSW,NSG,HM-ANN,DiskANN,KD-tree, R-Tree, Rstar-Tree, IVFADC, LSH, SymphonyQG, DEG, ELPIS, RkANNicde, ACORNsigmod, iQAN,wang2025accelerating,IP-DiskANN}, such as HNSW \cite{HNSW}, IVFADC \cite{IVFADC}, and Vamana \cite{DiskANN}. 
Most of these algorithms focus on offline constructing high-quality indexes for high precision and low-latency vector queries. 
However, only a few algorithms address the online construction and updating of index structures (including insertions and deletions), such as R-tree \cite{R-Tree,Rstar-Tree} and FreshVamana \cite{FreshDiskANN}. 
R-tree indexes spatial data using multi-level bounding rectangles and supports efficient insertion and deletion operations within nodes, enabling dynamic updates to the data. When data is inserted or deleted, R-tree adjusts the size of bounding rectangles and reorganizes nodes as necessary to maintain the efficiency and balance of the index structure. 
FreshVamana \cite{FreshDiskANN} is the first graph-based index to support insertions and deletions. It incrementally adjusts and optimizes the index based on updates to avoid the expensive overhead of rebuilding the graph index, while maintaining high index quality.

\stitle{Vector Search Systems.}
To meet the vector search performance requirements in different real-world scenarios, numerous vector search systems have been developed \cite{spfresh, FreshDiskANN, DiskANN, OOD-DiskANN, Filtered-DiskANN, BANG, SPTAG, KGraph, Starling, SPANN, Tribase, UNGsigmod25, SuCosigmod25, DIDS, ParlayANN, Auncel,Pan0L24}. These systems typically optimize vector search performance by combining specific index structures and search algorithm characteristics. Some work further explores combining disk and other external storage devices to support large-scale vector search under limited memory (e.g., \cite{DiskANN, SPANN}). 
However, most of these works focus on query performance optimization in static vector data indexing. 
Only a few systems support the dynamic update (insertion and deletion) of index structures in dynamic vector scenarios, such as SPFresh \cite{spfresh} and \freshdiskann \cite{FreshDiskANN}. 
As introduced in Section \ref{sec:bg:existing_work}, SPFresh \cite{spfresh} uses a cluster-based index and performs poorly in high-dimensional vector scenarios, while \freshdiskann provides limited update performance.

\stitle{Vector Database.}
To efficiently manage vector data and support complex query requirements, 
numerous vector databases have been developed \cite{AnalyticDB-V, PASE, SingleStore-V, pgvector, VBASE, Milvus, Pinecone, vexless, Vearch, Faiss}. 
These vector databases can generally be categorized into two types: generalized vector databases and specialized vector databases \cite{ZhangLW24}. 
Generalized vector databases integrate vector data management functionalities into existing relational database systems such as AnalyticDB-V \cite{AnalyticDB-V} and PASE \cite{PASE}. 
This allows them to reuse many features of the existing systems, such as distributed architecture, high availability guarantees, and SQL support. 
However, they often sacrifice query performance to some extent due to the limitations of traditional database components that hinder fine-grained optimization of vector data \cite{Milvus}. Specialized vector databases, 
are purpose-built data management systems specifically for vector data, designed to efficiently store and search large-scale vector data \cite{Milvus, Pinecone, vexless, Vearch, Faiss}. 
They treat vector data as a first-class citizen, which allows them to optimize system performance more effectively, typically achieving better performance.
However, both existing types of vector databases have functional or performance limitations when it comes to executing vector updates, particularly for deletions. 
They often require periodically rebuilding the entire index to apply updates \cite{spfresh}, which is costly.

%% file: 7-concl.tex
\section{Conclusion}
\label{sec:concl}

This paper proposes a topology-aware localized update strategy for graph-based ANN indexes that supports rapid index updates while maintaining high search efficiency and accuracy.
It employs three key designs to improve the system's update performance. 
Firstly, we propose a topology-aware index to quickly identify affected vertices by avoiding unnecessary I/O caused by reading large amounts of vector data.
Secondly, we design a localized update method, updating the index locally based only on the affected vertices, thereby avoiding modifications to the entire index file. 
Finally, we propose a similar neighbor replacement method that replaces deleted vertices with a small number of similar neighbors, thus avoiding frequently triggering expensive neighbor pruning.
Based on the above strategy, we implement a system, \oursys. The evaluation results confirm the effectiveness and efficiency of our system.